\documentclass[fleqn,10pt]{wlscirep}
\usepackage[utf8]{inputenc}
\usepackage[T1]{fontenc}
\usepackage{appendix}
\usepackage{caption}
\usepackage{ragged2e}
\usepackage{pifont}
\usepackage{fontawesome}
\usepackage[left]{lineno}
\usepackage{natbib}
\setcitestyle{authoryear,round,maxcitenames=2}
\usepackage{xcolor}
\usepackage{listings}
\usepackage{booktabs}
\usepackage{array}
\newcommand{\up}{$\uparrow$}
\newcommand{\dn}{$\downarrow$}

\definecolor{cbgBack}{HTML}{F7F8FA}   
\definecolor{cbgRule}{HTML}{E1E4E8}   
\definecolor{cbgText}{HTML}{24292E}   
\definecolor{cbgKey} {HTML}{0033B3}   
\definecolor{cbgStr} {HTML}{067D17}   
\definecolor{cbgCom} {HTML}{8C8C8C}   
\definecolor{cbgNum} {HTML}{1750EB}   
\definecolor{cbgFun} {HTML}{795E26}   
\definecolor{cbgLine}{HTML}{B0B5BD}   

\lstdefinestyle{cigpy}{%
	language=Python,
	basicstyle=\ttfamily\footnotesize\color{cbgText},
	keywordstyle=\color{cbgKey}\bfseries,
	stringstyle=\color{cbgStr},
	commentstyle=\color{cbgCom}\itshape,
	numberstyle=\ttfamily\scriptsize\color{cbgLine},
	emphstyle=\color{cbgFun},
	backgroundcolor=\color{cbgBack},
	frame=single,
	framesep=6pt,
	framerule=0.4pt,
	rulecolor=\color{cbgRule},
	xleftmargin=14pt,
	xrightmargin=4pt,
	numbers=left,
	numbersep=8pt,
	showstringspaces=false,
	keepspaces=true,
	columns=fullflexible,
	breaklines=true,
	breakatwhitespace=true,
	breakindent=12pt,
	postbreak=\mbox{\textcolor{cbgLine}{$\hookrightarrow$}\space},
	tabsize=4,
	upquote=true,
	aboveskip=8pt,
	belowskip=4pt,
	emph={FaultPredictor,RGTPredictor,ChannelPredictor,KarstPredictor,%
		PropertyPredictor,HRNet,HRNetSkipOpt,np},
}

\lstdefinestyle{cigshell}{%
	basicstyle=\ttfamily\footnotesize\color{cbgText},
	backgroundcolor=\color{cbgBack},
	commentstyle=\color{cbgCom}\itshape,
	frame=single,
	framesep=6pt,
	framerule=0.4pt,
	rulecolor=\color{cbgRule},
	xleftmargin=6pt,
	xrightmargin=4pt,
	showstringspaces=false,
	keepspaces=true,
	columns=fullflexible,
	breaklines=true,
	tabsize=4,
	aboveskip=8pt,
	belowskip=4pt,
	morecomment=[l]\#,
}

\title{Artificial Intelligence for Subsurface Imaging Understanding: A Decade Review of Challenges, Methods, Benchmarks, and Outlook}

\author[1]{Yimin Dou}
\author[1,*]{Xinming Wu}
\author[1]{Hui Gao}
\author[2]{Mingliang Liu}
\author[3]{Tao Zhao}
\author[4]{Zhi Zhong}
\author[3]{Haibin Di}
\author[5]{Min Jun Park}
\author[6]{Robert G. Clapp}
\author[1]{Zhixiang Guo}
\author[1]{Long Han}
\author[7]{Sergey Fomel}

\affil[1]{School of Earth and Space Sciences, University of Science and Technology of China, Hefei, China}
\affil[2]{School of Future Technology, Shandong University, Jinan, China}
\affil[3]{SLB, Houston, TX, USA}
\affil[4]{China University of Geosciences (Wuhan), Wuhan, China}
\affil[5]{X, the Moonshot Factory (Google X), Mountain View, CA, USA}
\affil[6]{Stanford University, Stanford, CA, USA}
\affil[7]{The University of Texas at Austin, Austin, TX, USA}

\affil[*]{Corresponding author: xinmwu@ustc.edu.cn \par \faGithub~\textbf{Homepage}: \url{https://cig.ustc.edu.cn/cigBench}   \par \url{https://douyimin.github.io/CIG-bench}}

\begin{abstract}

	Subsurface imaging understanding bridges observed geophysical data and quantitative geological models, supporting applications such as hydrocarbon exploration, CO$_2$ storage site assessment, and geohazard monitoring. Over the past decade, machine learning and, increasingly, deep learning have substantially reshaped interpretation workflows. This review synthesizes the literature from 2015 to 2025 across four major tasks, namely structural interpretation, geobody identification, seismic facies analysis, and property estimation, tracing how the field evolved from classical machine learning through deep learning to emerging domain foundation models, and how the four tasks are coupled within a single interpretation system.
	Subsurface imaging interpretation remains fundamentally different from other AI-driven tasks, facing ambiguous signals, pronounced interpretive non-uniqueness, sparse semantics, unfixed target locations, and scarce reliable annotations. Building on the reviewed literature, we synthesize three interrelated challenges that define its frontier: interpretation under complex geological conditions, cross-survey semantic generalization under low information density, and the absence of reliable benchmarks. We argue that addressing them will hinge on integrating human expertise, physical constraints, and geological priors into model training and inference, and on treating uncertainty quantification as an intrinsic model output. From this basis we outline a forward-looking agenda: unified, jointly modelled interpretation systems with explicit cross-task consistency; priors that evolve from physics toward language and multimodal supervision; end-to-end uncertainty propagation; human--AI collaboration and agent-orchestrated workflows; and a more rigorous evaluation science supported by an AI-ready data ecosystem.
	Among these challenges, the lack of unified benchmarks has been particularly consequential, making fair method comparison difficult, hindering reproducibility, and fragmenting community efforts. We review the state of benchmarking across the four tasks, clarify why an objective field ground truth is unattainable, and outline what reproducible, uncertainty-aware evaluation in this field would require. To make these arguments concrete and reproducible, the review is accompanied by an open, evolving benchmark resource (CIG-Bench)---covering fault segmentation, relative geologic time estimation, geobody segmentation, and property modeling with synthetic datasets, pretrained baselines, and a quantitative evaluation---used in the main text only for qualitative field examples, with its full construction and quantitative results provided as Supplementary Material and maintained at \url{https://douyimin.github.io/CIG-bench}. We hope this decade-spanning review offers a timely reference for the field and helps move subsurface imaging interpretation toward a more systematic and reproducible discipline.

\end{abstract}
\newpage
\begin{document}

	\flushbottom
	\maketitle
	\thispagestyle{empty}

	\par
	\newpage

	{
		\setlength{\parskip}{0pt}
		\section*{Contents}
		\vspace{-0.5em}
		\renewcommand{\contentsname}{}
		\setcounter{tocdepth}{2}
		\makeatletter
		\renewcommand*\l@section[2]{%
			\addpenalty\@secpenalty
			\addvspace{0.4em \@plus\p@}%
			\setlength\@tempdima{1.5em}%
			\begingroup
			\parindent \z@ \rightskip \@pnumwidth
			\parfillskip -\@pnumwidth
			\leavevmode \bfseries
			\advance\leftskip\@tempdima \hskip -\leftskip
			#1\nobreak\normalfont
			\leaders\hbox{$\m@th\mkern \@dotsep mu\hbox{.}\mkern \@dotsep mu$}\hfill
			\nobreak\hb@xt@\@pnumwidth{\hss\bfseries #2}\par
			\endgroup}
		\renewcommand*\l@subsection[2]{%
			\addpenalty\@secpenalty
			\addvspace{0pt}%
			\setlength\@tempdima{2.3em}%
			\begingroup
			\parindent \z@ \rightskip \@pnumwidth
			\parfillskip -\@pnumwidth
			\leavevmode
			\advance\leftskip\@tempdima \hskip -\leftskip
			#1\nobreak
			\leaders\hbox{$\m@th\mkern \@dotsep mu\hbox{.}\mkern \@dotsep mu$}\hfill
			\nobreak\hb@xt@\@pnumwidth{\hss #2}\par
			\endgroup}
		\makeatother
		\setlength{\parskip}{0pt}
		\linespread{0.95}\selectfont
		\tableofcontents
	}
	\vspace{1em}

	\section{Introduction}\label{sec:intro}

	\begin{figure*}[!htb]
		\includegraphics[scale=0.58]{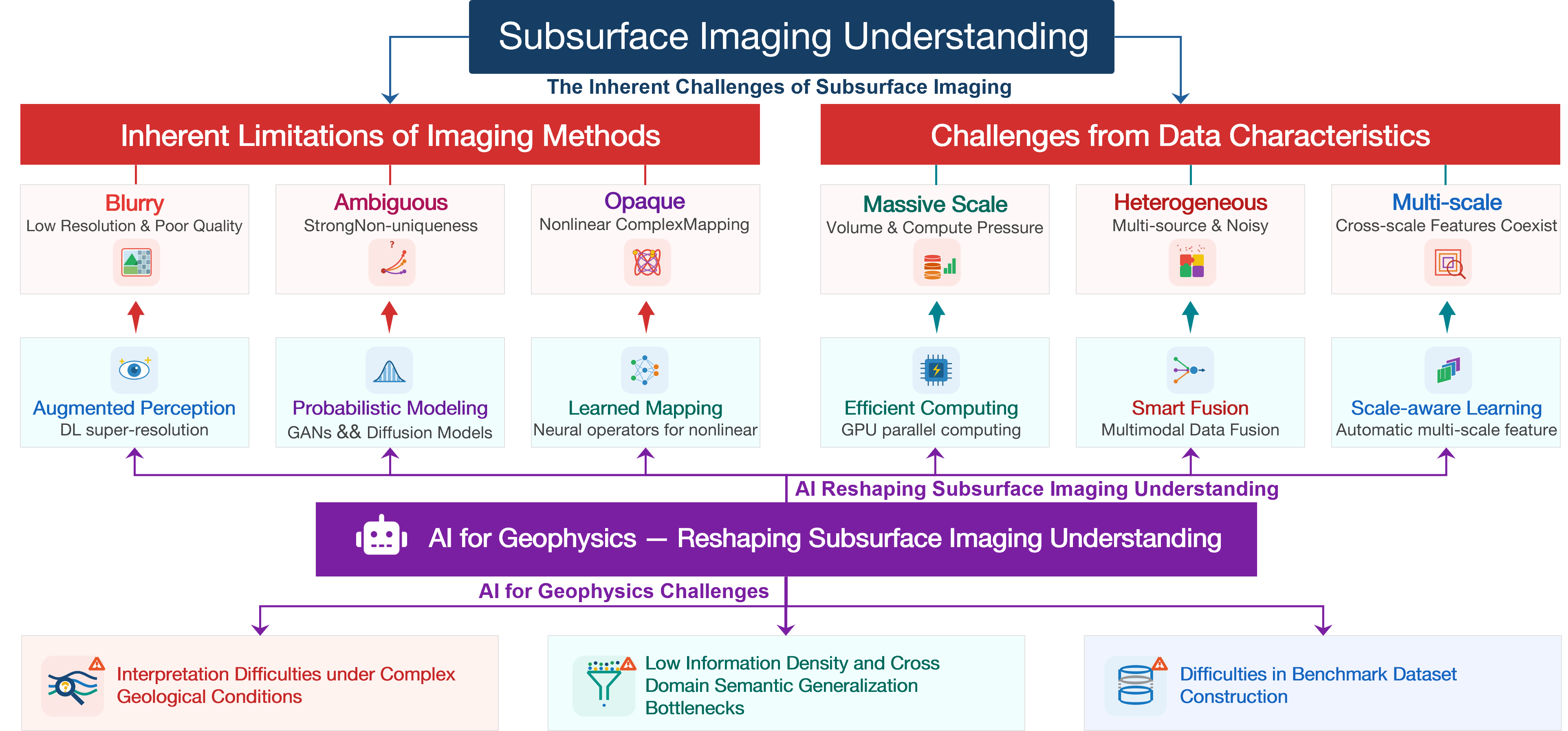}
		\centering\caption{ A structured overview of AI for geophysics-based subsurface imaging understanding. The framework identifies two categories of core challenges: inherent imaging limitations (blurriness, ambiguity, and nonlinear mapping) and data characteristics (massive scale, heterogeneity, and multi-scale features). Against these challenges, AI-driven solutions including deep learning-based perception, probabilistic modeling, physics-informed learned mapping, efficient computing, multimodal fusion, and scale-aware learning are shown to reshape existing research paradigms. The framework further highlights three open challenges for future research: interpretation under complex geological conditions, cross-survey semantic generalization, and benchmark dataset construction.
		}
		\label{fig00}
	\end{figure*}

	Human understanding of the Earth's interior began with indirect observations at the surface and information obtained through drilling. Yet the composition, structural characteristics, and dynamic processes of the deep subsurface have long remained central unresolved questions in Earth science. Unlike astronomical observations, which permit direct visualization of distant galaxies, probing the Earth's interior faces a fundamental limitation: geological structures and physical properties at depths of hundreds to thousands of meters cannot be directly observed. This intrinsic invisibility makes subsurface imaging an indispensable tool for revealing the hidden architecture and processes of the Earth \citep{mousavi2022deep,yu2021deep,bergen2019machine}.

	However, the technologies associated with subsurface imaging are inherently indirect and correspond to a fundamentally ill-posed inverse problem \citep{grana2016bayesian}. They aim to infer highly complex underground structures using only limited surface observations acquired through physical measurement systems. These observations are constrained by limited aperture, which restricts the range of illumination and observation; limited frequency bandwidth, which imposes resolution limits through wavelength; and noise contamination, which reduces signal fidelity. As a consequence of these physical constraints, seismic imaging inevitably exhibits restricted resolution, artifacts introduced by imperfect illumination and modeling, and significant non-uniqueness. These inherent deficiencies make the reliable reconstruction of the Earth's interior both challenging and uncertain.

	A systematic review of subsurface interpretation literature published between 2015 and 2025 reveals that machine learning and artificial intelligence methods are steadily becoming the dominant research paradigm, with their proportion in the literature increasing markedly each year \citep{bergen2019machine,yu2021deep}. This trend is driven primarily by two factors. First, machine learning has already transformed interpretation workflows in related domains such as medical image analysis and remote sensing, and its mature methodologies and toolchains provide a valuable reference for subsurface interpretation \citep{wang2018successful,khosroanjom2024machine}. Second, machine learning has demonstrated substantial performance gains in several key tasks within subsurface interpretation, including fault detection \citep{wu2019faultseg,xiong2018seismic}, horizon picking \citep{wu2019semiautomated,geng2020deep}, seismic facies classification \citep{alaudah2019machine,qian2018unsupervised}, and impedance inversion \citep{yang2019deep,das2019convolutional}, offering empirical evidence that it represents a practical and promising technical pathway for future development.

	Building upon machine learning, intelligent methods represented by deep learning offer significant advantages over traditional subsurface imaging and conventional machine learning approaches \citep{yu2021deep,mousavi2024applications}. First, deep networks enable end-to-end feature learning, automatically extracting multi-level representations from raw seismic data that range from textures to semantic patterns; convolutional and self-attention based modules function as task-adaptive "seismic attributes", thereby reducing reliance on hand-crafted feature design. Second, deep learning provides strong nonlinear mapping capability, allowing the approximation of high-dimensional functional relationships that arise from coupled physical processes, and capturing implicit patterns that are difficult for linear or weakly nonlinear methods to represent \citep{biswas2019prestack,sun2021physics,li2020deep}. Third, once trained, deep models achieve highly efficient parallel inference on GPUs or TPUs, enabling large-scale real-time or near-real-time interpretation; for tasks such as fault detection, the computational efficiency can exceed that of traditional workflows by several orders of magnitude. Fourth, deep models offer output consistency, producing stable results for identical inputs, which reduces interpreter subjectivity and facilitates multi-temporal monitoring and regional-scale comparative analysis. Fifth, these models possess continual learning capability, allowing iterative updates and performance improvement through incremental data accumulation, thereby creating a positive feedback cycle between data growth and model capability \citep{cunha2020seismic,yan2021improving}.

	These advantages rely on the foundation of large-scale models and high-quality datasets \citep{alaudah2019machine,chai2022open,wu2019faultseg}. However, compared with data modalities in remote sensing or medical imaging, subsurface seismic data exhibit greater complexity and variability in terms of acquisition geometry, illumination, noise, and geological diversity \citep{wu2023sensing,khosroanjom2024machine}. Consequently, although research advances in artificial intelligence have been considerable, only fault interpretation has produced solutions that approach scalable and stable deployment in engineering practice, supported by a rapidly growing ecosystem of 3D fault detection networks and evaluation studies \citep{wu2019faultseg,wu2019faultnet,an2023current}. For most other essential subsurface interpretation tasks, mature AI-based systems that support large-scale operational use have yet to be established. As the bridge linking seismic information with geological understanding, subsurface imaging and interpretation still face a wide range of systemic challenges.

	The field of subsurface imaging understanding confronts challenges from two fundamental dimensions, as illustrated in Figure~\ref{fig00}. The first stems from inherent limitations of imaging methods: seismic data are inherently blurry due to limited bandwidth and aperture, ambiguous due to strong non-uniqueness where identical observations may correspond to multiple geological scenarios, and opaque due to the nonlinear complex mapping between surface measurements and subsurface properties. The second dimension arises from data characteristics, including the massive scale of terabyte-level 3D surveys, heterogeneity introduced by multi-source noisy acquisition, and multi-scale complexity where features ranging from meter-scale fractures to kilometer-scale salt bodies must be simultaneously handled. In response to these challenges, AI has driven a series of technical advances over the past decade, with augmented perception addressing blurriness, probabilistic modeling tackling ambiguity, learned mapping resolving opacity, efficient computing managing massive scale, smart fusion handling heterogeneity, and scale-aware learning capturing multi-scale structures. Together, these approaches represent a paradigm shift from fragmented task-specific models toward unified, transferable frameworks. Despite this progress, three open challenges persist and define the frontier of current research: the difficulty of interpreting complex geological conditions, cross-survey semantic generalization bottlenecks arising from low information density, and the lack of reliable benchmark datasets. The last of these directly motivates the open benchmark resource (CIG-Bench) that accompanies this review and that we use as a concrete, reproducible illustration of the field-wide arguments developed below.

	\subsection{AI for Geophysics Challenges}\label{sec:challenges}

	\begin{figure*}[!htb]
		\includegraphics[scale=0.5]{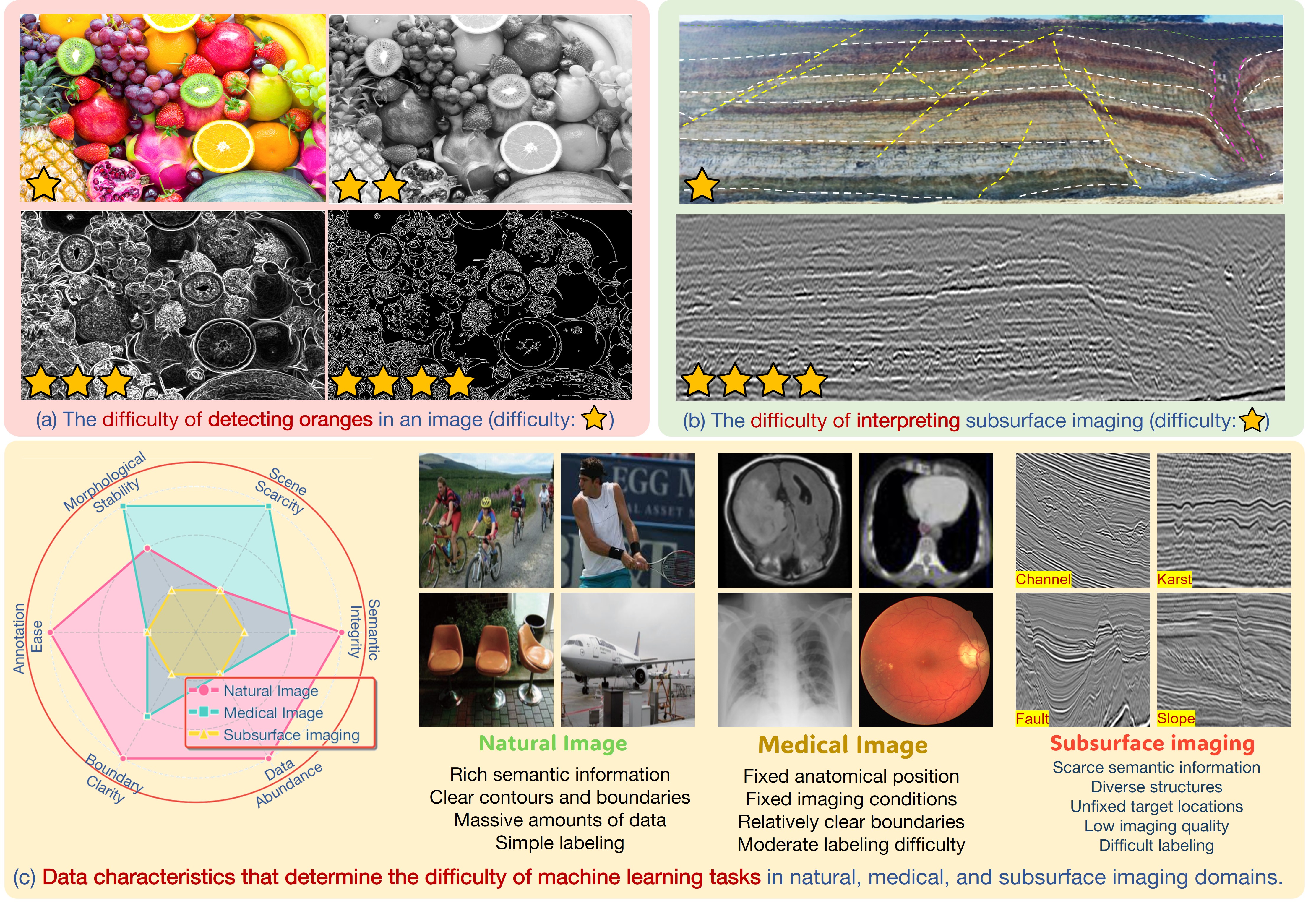}
		\centering\caption{Adapted from \cite{yimindou2025geological}. (a, b) A comparison of task difficulty between orange detection in natural images and subsurface interpretation. Natural images offer rich visual cues, whereas subsurface targets must be inferred from weak and ambiguous seismic responses. (c) A comparison of natural, medical, and subsurface imaging. In contrast to natural and medical images, subsurface imaging has sparse semantics, diverse structures, low image quality, and difficult annotation, making interpretation and generalization substantially more challenging.
		}
		\label{fig0}
	\end{figure*}

	Compared with natural images and medical images, subsurface imaging data present substantially greater challenges for machine learning tasks. As illustrated in Figure~\ref{fig0}, natural images benefit from rich semantic information, clear contours and boundaries, and abundant annotated data, while medical images enjoy relatively well-defined target boundaries owing to fixed anatomical positions and standardized imaging conditions. Subsurface imaging data, by contrast, are characterized by scarce semantic information, diverse structural morphologies, unfixed target locations, low imaging quality, and considerable labeling difficulty. The radar chart further reveals that subsurface imaging lags far behind the other two modalities in terms of data abundance, boundary clarity, and semantic integrity—intrinsic properties that collectively determine the high difficulty of intelligent interpretation tasks. These challenges can be understood from three interrelated perspectives: interpretation difficulties under complex geological conditions, bottlenecks in low information density and cross-survey semantic generalization, and inherent limitations in benchmark dataset construction.

	\subsubsection{Interpretation Difficulties under Complex Geological Conditions}

	Interpretation challenges associated with complex geological settings arise from multiple sources \citep{yu2021deep,mousavi2022deep}. At the structural system level, complexity is driven by multi-phase tectonic overprinting. En echelon, imbricated, and flower like fault systems exhibit intricate geometric relationships, and small displacement faults are easily missed due to their weak seismic expression. In the shallow subsurface, scattering and absorption within fractured zones combined with velocity heterogeneity lead to energy shadow zones and traveltime distortions in underlying layers \citep{feng2021uncertainty,an2021deep}. In the deep subsurface, resolution loss causes faults to appear blurred, discontinuous, or misaligned, which complicates the reconstruction of continuous structural features \citep{li2024faultseg,dou2024faultssl}.

	Special tectonic environments introduce additional difficulties \citep{adler2021deep,sun2021physics}. Subduction zone megathrusts exhibit sharp velocity contrasts and strong anisotropy, which combined with multiple wave interference make interface tracking particularly challenging. Strike slip fault zones contain abrupt lateral velocity changes that induce striped artifacts and false anomalies, thereby obscuring true structures. Fold–unconformity systems display high uncertainty in stratigraphic relationships \citep{bi2021deep,geng2020deep}. Intense deformation produces abrupt variations in dip and highly dispersed reflection patterns, while truncation and wedge out at unconformities complicate sequence correlation. Strong velocity anomaly bodies such as salt bodies and igneous intrusions further amplify interpretational uncertainty \citep{waldeland2018convolutional,wu2016methods}. Their wavefield effects generate shadow zones, scattering induced "smile" artifacts, and false fault responses, which increase ambiguity in fault connectivity and displacement estimation and weaken the reliability of three-dimensional interpretation \citep{wang2015noise,alfarhan2022robust}.

	\subsubsection{Low Information Density and Cross-Survey Semantic Generalization Bottlenecks}

	The difficulty of cross-survey generalization in seismic interpretation arises from three systemic levels. At the data level, heterogeneity in geological settings leads to large variations in geometry, physical contrast, and spatial scale of target bodies \citep{cunha2020seismic,nasim2022seismic}. Variations in acquisition parameters cause discrepancies in frequency content, signal-to-noise ratio, and azimuthal anisotropy. Non-standardized processing workflows reshape statistical characteristics and texture patterns. At the physical level, the indirect and ill-posed nature of seismic imaging inherently produces low information density and weak semantic content. At the interpretation level, these characteristics lead to severe non-uniqueness; identical seismic responses may correspond to multiple geological scenarios, which is particularly critical in seismic facies analysis and property inversion \citep{grana2022probabilistic,feng2021uncertainty,qian2018unsupervised}.

	Limitations of purely data-driven models exacerbate this challenge \citep{khosroanjom2024machine,li2023comprehensive,wu2023sensing}. Without the integration of physical constraints or geological priors, such models tend to overfit distributional characteristics of the training domain and exhibit semantic drift when applied to new surveys.

	Viewed from the perspective of imaging characteristics, seismic data exhibit a "dual disadvantage" \citep{yimindou2025geological}. Their information density is lower than that of natural images, which means that informative features are much sparser. Their structural complexity is higher than that of medical images, which benefit from relatively fixed imaging geometries and anatomical priors, whereas geological structures show greater uncertainty and variability in spatial morphology, scale hierarchy, and structural combinations. The combination of information sparsity and structural complexity imposes a dual challenge for the generalization capability and geological consistency of intelligent interpretation models.

	Empirical evidence shows that current methods still suffer from significant limitations in cross-survey generalization. Most models require survey-specific fine tuning to achieve acceptable accuracy in new areas. Only in relatively stable scenarios, such as shallow fault detection, do models exhibit some plug and play capability. A genuinely universal modeling paradigm for subsurface interpretation has not yet been established \citep{sheng2025seismic,yimindou2025geological,gao2026foundation}.

	\subsubsection{Difficulties in Benchmark Dataset Construction}

	Data represent a fundamental resource for artificial intelligence and form the essential infrastructure for advancing subsurface imaging and interpretation \citep{bergen2019machine,alaudah2019machine}. A systematic benchmark unifies task definitions, labeling standards, and evaluation protocols, and provides a reproducible experimental environment that supports fair performance comparison and methodological iteration. Such a benchmark also offers a reliable foundation for model training, accelerates model development, reduces redundant research cost, and enables standardized collaboration across institutions.

	Two technical pathways have been explored, yet both face intrinsic limitations. The path based on manual annotation of real seismic data is constrained by the inherent non-uniqueness, ambiguity, and low semantic content of seismic reflections. Even experienced interpreters struggle to achieve high consistency when working on the same seismic volume \citep{wu2019semiautomated,zhang2021automatic}. Current workflows rely mainly on two-dimensional section based labeling, which cannot ensure geometric continuity or topological consistency across slices \citep{dou2022attention,dou2022loss}. Large scale three-dimensional labeling is therefore constrained by both cost and feasibility.

	The synthetic data generation path is limited by the extreme complexity and diversity of geological structures and depositional processes \citep{wu2019faultseg}. Existing synthetic modeling schemes cannot fully cover the combinatorial space of structural styles, sedimentary patterns, imaging artifacts, and noise conditions present in real surveys. This leads to a synthetic to real gap, which reduces transferability and weakens generalization performance of models trained purely on synthetic data. In recent years, several large-scale benchmark datasets have been released to address this challenge, including cigFacies for seismic facies classification \citep{gao2025cigfacies} and cigChannel for 3D channel segmentation \citep{wang2025cigchannel}, which provide standardized training and evaluation resources for the community.

	\subsection{Tasks}

	Subsurface imaging interpretation is a complex problem that targets multiple tasks and requires the joint incorporation of diverse geological and physical constraints. Considering the heterogeneity of interpretation objects and task objectives, this review categorizes subsurface interpretation into four major classes: structural interpretation, geobody interpretation, seismic facies interpretation, and property interpretation. Each class corresponds to specific geological meanings and exhibits distinct characteristics in terms of technical pathways and application scenarios. The definitions adopted in this paper are intended only for the narrow context of subsurface interpretation and should not be regarded as general definitions of these concepts in broader settings.

	\textbf{Structural interpretation:} Structural interpretation refers to the identification, characterization, and representation of subsurface geological structural elements. It encompasses, but is not limited to, the interpretation of key structural interfaces such as faults, horizons, and unconformities, and also includes related tasks such as constructing structural frameworks, estimating relative geologic time (RGT), and geological structural modeling.

	\textbf{Geobody interpretation:} Geobody interpretation aims to identify, characterize, and segment three-dimensional entities in the subsurface that exhibit relatively independent geometries and distinct geological origins. It covers typical geobody types such as igneous intrusions and traps, including but not limited to salt bodies, karst caves, and channels.

	\textbf{Seismic facies interpretation:} Seismic facies refers to a set of seismic reflection units exhibiting similar reflection characteristics, such as amplitude, frequency, continuity, waveform, and stratification patterns, and it typically reflects specific depositional environments or geological processes. At present, seismic facies interpretation is mainly conducted under two viewing paradigms: sectional views (also referred to as profile-based views, performed on inline/crossline sections) and horizon-based views (also referred to as the stratal perspective, performed along horizon-aligned or stratigraphically constrained slices).

	\textbf{Property interpretation:} Here, "properties" denote subsurface geological or geophysical parameters — including elastic attributes such as impedance and velocity, and petrophysical quantities such as porosity and density — that are inferred from seismic data and calibrated through well logs. The goal of property interpretation is to establish generalizable mappings between seismic responses and physical parameters, enabling the inversion, estimation, and modeling of property volumes.

	\subsection{Contribution}
	From a macro perspective, Bergen et al. systematically discuss the applications of machine learning in solid Earth sciences and emphasize the critical role of open science principles and benchmark datasets in sustaining long-term disciplinary development \citep{bergen2019machine}. Mousavi and Beroza focus on the use of deep learning across subfields of seismology, covering earthquake signal detection, denoising, and image-based interpretation, among other topics \citep{mousavi2022deep}. Yu and Ma provide a systematic overview of recent advances and future trends of deep learning in geophysics, with coverage that spans exploration geophysics, seismology, and atmospheric sciences. These reviews offer researchers a robust macroscopic perspective and a clear conceptual framework, but they primarily target the broader Earth science community \citep{yu2021deep}. Their discussions remain relatively shallow with respect to specific tasks in subsurface imaging and interpretation, and therefore provide limited detailed and operational guidance for method selection and technical implementation on particular interpretation problems.

	For specific interpretation tasks in subsurface imaging, several high-quality task-oriented reviews are already available. An et al. concentrate on fault interpretation and provide a detailed synthesis of deep-learning model architectures, datasets, and evaluation metrics; their work is one of the most systematic and comprehensive reviews currently available for fault detection \citep{an2023current}. Xu et al. review the development of seismic facies analysis from a geological perspective and highlight the importance of integrated geological and geophysical constraints \citep{xu2022seismic}. Islam et al. focus on the identification of salt bodies and summarize deep-learning-based salt segmentation methods together with related public datasets \citep{islam2024comprehensive}. Li et al. present a comprehensive review of the applications of fully connected networks, convolutional networks, recurrent networks, and generative networks in seismic inversion \citep{li2023comprehensive}, while Wang et al. further summarize recent advances of deep neural networks in velocity modeling and impedance inversion \citep{wang2023deepa}. Grana et al. concentrate on probabilistic inversion methods and emphasize the central role of uncertainty quantification in reservoir characterization \citep{grana2022probabilistic}. Overall, these single-task reviews have reached a high level of technical depth within their respective domains. However, they are often restricted to a particular class of geological bodies or a single type of structure and have not yet examined, from an integrated perspective, the unified framework, benchmark construction, and coupled evolution of the entire subsurface interpretation task system in the era of deep learning and large models.

	In the domain of exploration geophysics and subsurface imaging, a number of topic specific reviews have also been published. Mousavi et al. systematically summarize the use of deep neural networks throughout the seismic exploration workflow, including data acquisition, preprocessing, migration imaging, and interpretation, thereby providing a clear technical roadmap for constructing an end-to-end intelligent seismic exploration workflow \citep{mousavi2024applications}. Khosro Anjom et al. examine the current status and limitations of machine learning in seismic exploration from a more critical standpoint, construct a systematic literature database, and perform quantitative analyses of research hotspots and their temporal evolution \citep{khosroanjom2024machine}. Wu et al. summarize three major strategies for incorporating domain knowledge into deep neural networks, namely the use of geological and geophysical forward modeling to generate synthetic training data, the explicit embedding of physical operators and preconditioners into network architectures, and the incorporation of mechanical and geological constraints into loss functions, with the goal of achieving an organic integration of data-driven and physics-driven paradigms \citep{wu2023sensing}. Lin and Zhong et al. systematically review the application of machine learning to the interpretation of faults, salt bodies, horizons, channels, and karst systems, and they release corresponding annotated datasets, which provide an important foundation for subsequent model evaluation and method comparison \citep{lin2024machine}. Taken together, these studies promote the dissemination and standardization of machine learning in seismic exploration from the perspectives of exploration workflow, methodology, and application case studies. Nevertheless, they have not yet developed a systematic discussion that uses the subsurface imaging interpretation task system as the central object. The internal connections among sub tasks such as structural interpretation, geobody delineation, seismic facies analysis, and property interpretation, as well as a unified evaluation framework for these tasks, remain relatively under explored. A fully developed, task oriented solution framework and a reproducible benchmark system dedicated to subsurface interpretation have also yet to emerge.

	The main contributions of this review are as follows.

	\begin{itemize}
		\item \textbf{A decade-spanning, task-coupled review.} We synthesize the 2015--2025 literature across four interpretation tasks---structural interpretation, geobody identification, seismic facies analysis, and property estimation. Rather than treating them in isolation, we trace their internal connections and co-evolution, assess how deep learning and AI have reshaped each task, and evaluate the strengths and limitations of current methods.

		\item \textbf{A synthesis of field-wide challenges and future directions.} From the reviewed literature we distill three durable, cross-task challenges---interpretation under complex geological conditions, cross-survey semantic generalization under low information density, and the absence of reliable benchmarks---and argue, with supporting evidence from the task-wise reviews, for the integration of physical and geological priors and for treating uncertainty as a first-class model output.

		\item \textbf{An accompanying open benchmark resource.} To make the above discussion concrete and reproducible, the review is accompanied by CIG-Bench, an open and evolving resource providing synthetic datasets, pretrained baselines, and one-click inference APIs for fault segmentation, relative geologic time estimation, geobody segmentation, and property modeling. In the main text we use it only to provide qualitative field examples; its full construction, baseline implementations, evaluation protocol, and quantitative results are provided as Supplementary Material.
	\end{itemize}

	A quantitative, in-corpus overview of the curated literature---including the
	citation-network view across the four task categories and the publication-trend
	analysis---is provided in the Supplementary Material (Section~\ref{si:litpart}), in line
	with the journal's policy on bibliometric content. The discussion below is
	organised qualitatively around the four tasks.

	\section{Task-wise Review}\label{sec:taskreview}
	This chapter provides a systematic review of the past decade of machine learning and AI advances across four subsurface imaging and interpretation tasks: Structure, Geobody, Facies, and Property. Across the corpus reviewed here, the early stage (2015 to 2017) featured relatively few studies and slow growth; since 2018 the field entered a rapid expansion phase and remained at a high level after 2019, reaching a local peak around 2023, followed by a modest decline. Research effort has been unevenly distributed across the four tasks, with Property and Structure attracting the most attention and Facies and Geobody comparatively less; a quantitative breakdown of this distribution is provided in the Supplementary Material (Section~\ref{si:biblio}).

	This imbalance reflects shared priorities in both research and practice. Property modeling typically serves high-value applications such as reservoir characterization and quantitative prediction, generating direct benefits for reserve evaluation, development-plan optimization, and risk control, and has therefore remained the most engineering-driven and economically impactful direction. Structure, in contrast, focuses on recovering fault systems and stratigraphic frameworks; its outputs provide essential geometric constraints for horizon tracking, geobody boundary delineation, facies-belt mapping, and the spatial consistency of property modeling, and often determine the upper bound of reliability for subsequent interpretation, making it a foundational component across multiple subsurface tasks.

	By comparison, Facies and Geobody more often function as intermediate representations that bridge seismic responses and geological semantics. They are typically used within an established structural framework to support depositional-unit delineation, facies zoning, and object-based characterization, and they provide priors for property modeling, including lithologic assemblages, depositional environments, and reservoir-body geometries. In many industrial workflows, they may also appear as derived products obtained after property inversion and reservoir prediction, for example through thresholding, clustering, or connectivity analysis. Because label definitions for these tasks are more subjective and scale-dependent, and are strongly influenced by local geological context, data quality, and imaging resolution, high-quality 3D annotation is costly and cross-survey generalization is difficult---factors that have constrained their research scale.

	Methodologically, the decade shows a clear transition: deep learning became dominant rapidly after 2018--2019, while the share of traditional machine learning continued to shrink, although it remains present in certain classification and feature-engineering-oriented studies. This shift provides the foundation for subsequent developments toward cross-survey generalization, stronger global-consistency constraints, and more interactive interpretation workflows.

	Throughout the task-wise reviews below, qualitative field examples produced with the accompanying CIG-Bench baselines are used purely to illustrate current capabilities and remaining failure modes. Because subsurface interpretation has no objective ground truth and densely annotated multi-survey field data essentially do not exist, the corresponding quantitative evaluation can only be conducted on synthetic data; that evaluation, together with its scope and caveats, is reported in the Supplementary Material (Section~\ref{si:cigpart}) and should be read as a controlled measure of relative ranking rather than a direct estimate of field performance.

	\subsection{Structure: Fault and Stratigraphic Framework Interpretation}
	\begin{figure*}[htb]
		\includegraphics[scale=0.64]{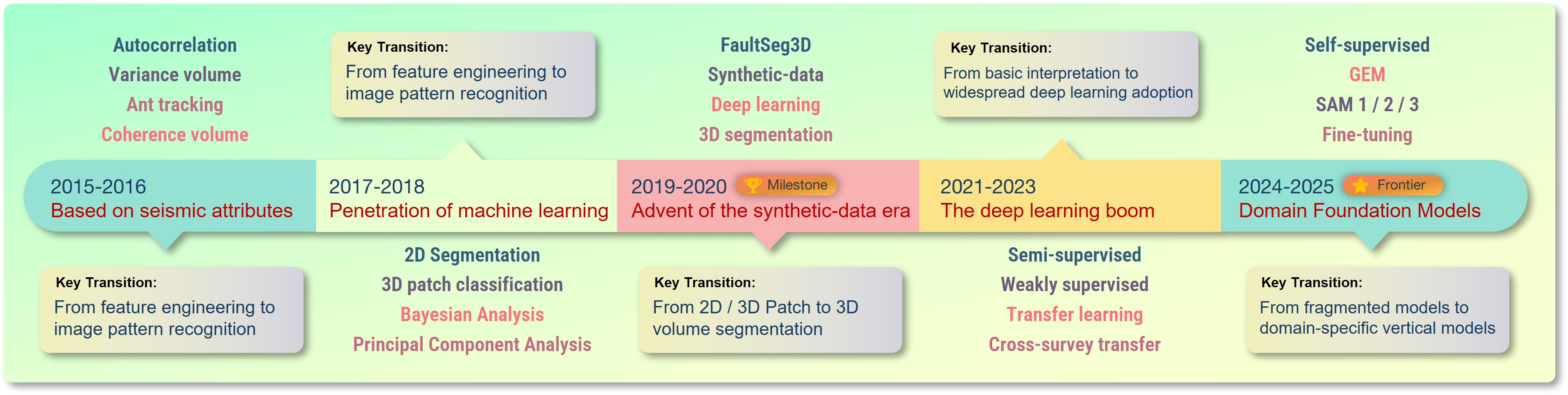}
		\centering\caption{The figure summarizes the technological evolution and paradigm shifts in structural interpretation over the past decade, driven by advances in machine learning and deep learning. Using stage-specific keywords and representative milestones, it delineates the research focus of each phase and its central transitions.
		}
		\label{fig2}
	\end{figure*}

	As shown in Figure~\ref{fig2}, progress in structural interpretation over the past decade can be summarized into five major phases, tracing an evolution from conventional seismic attribute analysis to deep learning and, more recently, to domain foundation models. The figure highlights the key technical transitions, representative keywords, and milestone events of each phase; we discuss them in turn in the following subsections. Whether the most recent move toward domain foundation models develops into a lasting paradigm shift, rather than a transfer of methods successful in other fields, still awaits longer-term validation in practical fault interpretation.

	\subsubsection{2015--2016: Seismic-attribute-driven interpretation}

	During 2015--2016, fault and horizon interpretation largely followed the classical paradigm of geometric modelling with attribute enhancement, where algorithms primarily improved visualisation and picking efficiency under interpreter control.

	For faults, representative efforts combined minimum/maximum autocorrelation factors with fuzzy classification to separate faulted from non-faulted zones in a high-dimensional attribute space \citep{shakiba2015fault}, alongside diffraction-separation and visual-saliency schemes built on coherence, curvature, and dip attributes. In practice, coherence, variance, and ant-tracking volumes remained the workhorses in industry.

	Geomechanical modelling was also employed to reduce structural uncertainty: \citet{maerten2015method} introduced geomechanically based restoration to constrain interpretations of poorly imaged horizons and faults, using stress-field consistency as an external validation.

	For horizons, \citet{herron2015pitfalls} catalogued common failure modes in autopicking and provided practical quality-control guidance, while \citet{labrunye2015merging} coupled global horizon tracking with relative geologic time modelling, merging extracted horizon patches into a space--time stratigraphic framework---an early use of RGT for stratigraphic correlation.

	Overall, structural interpretation in this stage remained interpreter-guided and attribute-centric, relying on refined attributes, geometric or geomechanical constraints, and improved tracking rather than data-driven feature learning, foreshadowing the deep-learning transition that followed.

	\subsubsection{2017--2018: Early adoption of machine learning}

	During 2017--2018, statistical machine learning and early deep learning began to systematically enter fault and horizon interpretation, while attribute-based methods continued to advance in parallel \citep{arayapolo2017automated,xiong2018seismic}.

	A first line of work refined geometric attributes. Directional structure-tensor coherence used derivatives taken parallel and perpendicular to local reflectors, rather than vertical/horizontal gradients, to highlight faults and subtle stratigraphic features better than eigenvalue-based coherence \citep{wu2017directional}; 3D flexure analysis, which measures the third-order variation of reflector geometry, delineated subtle faults and fractures overlooked by conventional curvature and discontinuity attributes \citep{di2017seismic}. Multi-attribute fusion using PCA further produced composite attributes for small-displacement faults \citep{jahan2017faulta}.

	A second line explicitly posed fault detection as supervised classification. \citet{guitton2017statistical} trained statistical models on labelled samples to predict 3D fault-probability volumes, among the earliest systematic machine-learning applications to fault imaging. \citet{arayapolo2017automated} learned a direct mapping from the seismic data space to fault locations, deliberately bypassing the computationally expensive physical-modelling and attribute-extraction steps; a Wasserstein-type loss was used so that predictions respect the structured, continuous nature of fault surfaces.

	From 2018 onward, CNN-based fault recognition expanded rapidly \citep{xiong2018seismic,wu2018convolutional,di2018seismic}. These studies treated 2D sections or 3D patches as images, enabling networks to learn fault textures and amplitude patterns and output pixel-level masks or probability volumes. \citet{xiong2018seismic} trained a CNN on annotated 3D field image cubes in a two-stage train-then-predict workflow that emulates how interpreters connect fault planes, showing that CNNs can match or exceed attribute-based methods; GAN-based augmentation was also introduced to enrich and diversify training data \citep{lu2018generative}. Subsequent workflows combined deep models with post-processing to enable end-to-end fault detection \citep{zhao2018fault,guo2018method,ma2018deep}.

	For horizons, multi-attribute and statistical strategies were incorporated into autopicking. \citet{lou2018automatic} fused phase, amplitude, and dip attributes to improve picking in complex settings, and hybrid frameworks combined texture segmentation, unsupervised clustering, and dynamic time warping for joint facies classification and horizon tracking \citep{bugge2018automatic}. However, deep-learning-based horizon interpretation remained comparatively limited.

	Overall, this period marked a transition from attribute engineering with shallow models toward data-driven pattern recognition. The adoption of deep learning was highly asymmetric: CNN-based fault detection surged, whereas horizon interpretation largely remained multi-attribute and statistical. Methods were still dominated by 2D CNNs on local patches, foreshadowing the move to fully 3D architectures in the next stage.

	\subsubsection{2019--2020: The rise of synthetic-data training}

	From 2019 to 2020, structural interpretation entered a fully deep-learning-driven stage in which 3D convolution and encoder--decoder architectures became central to fault and horizon workflows \citep{wu2019faultnet}. A milestone was FaultSeg3D \citep{wu2019faultseg}, which recast fault detection as binary 3D image segmentation and trained a fully convolutional (U-Net) network on roughly 200 automatically generated synthetic seismic volumes with paired fault labels, using a class-balanced loss to counter the extreme fault/non-fault imbalance. This relatively small synthetic set proved sufficient to train a network that generalised strongly across multiple field surveys, substantially alleviating the 3D annotation bottleneck and establishing the synthetic-data-driven paradigm for deep 3D fault segmentation. \citet{cunha2020seismic} then showed that a CNN pretrained on synthetic data could be transferred to a new field survey (the Netherlands F3 block) from as little as a single interpreted section, while \citet{pochet2019seismic} confirmed that synthetic poststack amplitude maps alone can serve as primary training data.

	Building on this paradigm, many works adopted encoder--decoder networks for voxel-level fault prediction on seismic volumes \citep{liu2020common}. Some frameworks jointly predicted fault probability along with strike and dip, enabling integrated estimation of geometric attributes \citep{wu2019faultnet}. Multi-task CNNs combined fault detection with structure-oriented smoothing and seismic-normal estimation to enhance fault responses while suppressing noise and acquisition footprints \citep{wu2019multitask}. UNet++ variants improved feature reuse \citep{yang2020seismic}, and structure-sensitive ``super-attributes'' were proposed to better emphasise small-offset faults and weak-reflection fractures as network inputs \citep{di2019improving}. Interpretability-guided CNNs also emerged to partially address the ``black-box'' concern by visualising seismic cues that drive segmentation decisions \citep{liu2020interpretability}.

	Horizon interpretation was similarly reshaped. Encoder--decoder CNNs enabled semi-automatic horizon extraction by predicting horizon images or probability volumes from 2D sections or 3D slices with minimal manual input \citep{wu2019semiautomated,tschannen2020extractinga,peters2019multiresolution}. Deep autoencoders were used to learn waveform embeddings for unsupervised horizon picking \citep{shi2020waveform}, and deep-learning-based RGT estimation provided a unified temporal framework for extracting stratigraphic sequences and horizons \citep{geng2020deep}. Semi-supervised stratigraphy workflows further combined sparse annotations with CNN training, including joint fault-and-stratigraphy interpretation on field data \citep{di2020seismic,di2020accelerating}.

	Early uncertainty quantification was also explored, including Bayesian deep-prior approaches for imaging and horizon tracking with uncertainty estimates and probabilistic tracking guided by transdimensional MCMC, as well as metrics that account for label ambiguity in fault and horizon evaluation \citep{siahkoohi2020uncertainty,cho2020semi,guillon2020ground}. Overall, synthetic data enabled a decisive shift from 2D slice-based workflows to full 3D volume segmentation, and the convergence of fault, horizon, and related tasks around 3D semantic-segmentation formulations laid the groundwork for subsequent advances.

	\subsubsection{2021--2023: Maturation of deep learning}

	Following the synthetic-data breakthrough, structural interpretation saw rapid expansion of deep-learning methods, shifting the focus from feasibility to systematic improvements in architectures, losses, learning paradigms, and interpretability \citep{an2021deep,gao2022automatic,wei2022seismic}. Fault and horizon interpretation were widely reformulated as voxel-level semantic-segmentation tasks, with architectures spanning fully convolutional networks, residual and nested U-Nets, multiscale attention CNNs, and 2.5D designs that trade computational cost for volumetric context \citep{wu2021fault,lin2022automatic}. For instance, the multiscale attention CNN (MACNN) merges and refines encoder feature maps across spatial resolutions through a spatial--channel attention mechanism \citep{gao2022automatic}, while a nested residual U-Net fuses low-, medium-, and high-resolution fault maps to produce clearer predictions on noisy images \citep{gao2022fault}. Transformer-based variants also emerged to capture longer-range dependencies, including 2.5D Transformer U-Nets and Transformer-assisted dual U-Nets, alongside recurrent CNNs with compound losses \citep{tang2023fault,wang2023transformer,ma2023seismic}.

	Loss-function design became a major emphasis due to extreme class imbalance, as faults often occupy less than 2\% of a seismic volume. Studies systematically combined BCE-style losses with overlap-based terms such as Dice and IoU, and explored multi-scale fusion with imbalanced-learning strategies \citep{wei2022seismic,li2023fault}. The MD loss was proposed to improve robustness under sparse or imperfect annotations by down-weighting anomalous labels \citep{dou2022loss}.

	Given the cost of 3D labels, weakly supervised and semi-supervised learning gained prominence. A complementary response to the synthetic-only paradigm was to release large expert-labelled field data: \citet{an2021deep} open-sourced a multi-gigabyte, expert-interpreted 3D field fault dataset and framed field fault recognition as an image-segmentation/edge-detection problem, providing a realistic benchmark beyond synthetic training. Attention-based 3D fault networks were trained with sparse 2D slice labels while leveraging large unlabelled volumes \citep{dou2022attention}. Knowledge distillation and structural augmentation using synthetic geometries provided additional pathways to improve label efficiency and expand training diversity \citep{wang2022distilling,wang2022structural}. Transfer learning and cross-survey adaptation were increasingly emphasised, typically pretraining on large synthetic datasets and fine-tuning on limited field labels, with progressive strategies designed to bridge differences in noise, bandwidth, and structural styles \citep{yan2021improving,zhou2021learning}. More physically realistic synthetic data generation, such as point-spread-function convolution, further improved field-data generalisation \citep{jing2023fault}.

	Research also began to incorporate geological reasoning explicitly. Interpretational constraints and human-reasoning principles were embedded into CNN workflows to improve geological plausibility, while reviews consolidated the state of the field and highlighted open directions \citep{di2021imposing,zhu2022fault,an2023current}. Uncertainty quantification matured as well: \citet{feng2021uncertainty} used a dropout-based approximation of Bayesian inference to decompose fault-prediction uncertainty into aleatoric and epistemic components, producing calibrated confidence maps alongside fault probabilities on the Netherlands F3 data and improving suitability for production workflows.

	A notable trend was the rise of DeepRGT methods: \citet{bi2021deep} trained a volume-to-volume, attention-equipped U-shaped network to regress a relative geologic time (RGT) volume from a seismic volume, from which 3D horizons and faults can be derived simultaneously, shifting horizon work from local geometric tracking toward global deep temporal modelling. Multi-task designs jointly predicted RGT, horizons, and faults using priors and Transformer components, and weakly supervised or sequence-constrained horizon tracking further improved label efficiency and stratigraphic consistency \citep{yang2023multi,wu2022variable,luo2023sequence}. Overall, 2021--2023 established deep learning as the dominant methodology for modern structural interpretation through coordinated advances in network design, losses, label efficiency, interpretability, and uncertainty estimation.

	\subsubsection{2024--2025: Emergence of domain foundation models}

	During 2024--2025, structural interpretation moved toward self-supervised pretraining, SAM-based prompt-driven architectures, and emerging domain foundation models, alongside continued advances in lightweight networks, label-efficient learning, and uncertainty quantification \citep{li2024faultseg,dou2024faultssl}.

	A defining trend is self-supervised pretraining for fault detection. \citet{zhang2024improving} introduced a 3D Transformer with multi-scale decoding, pretrained on unlabelled seismic data and fine-tuned for fault recognition, while contrastive and reconstruction-based objectives were explored to learn structural features without annotations \citep{dou2024seismic}. Pretrained 3D Transformers were also shown to support multiple downstream tasks, including fault detection, denoising, and horizon interpretation within a shared encoder \citep{wang2024trained}. Semi-supervised approaches such as FaultSSL combined limited labels with large unlabelled volumes to improve generalisation \citep{dou2024faultssl}.

	SAM-based architectures were adapted for interactive structural interpretation. Seismic Fault SAM enabled prompt-driven fault segmentation via lightweight modules and a 2.5D strategy, supporting point- or box-guided interaction \citep{chen2024seismica}. More broadly, multi-modal prompt engines and unified architectures were proposed for cross-survey segmentation of faults and other geobodies \citep{gao2026foundation}, and the promptable paradigm was further extended to multiple subsurface tasks within a single foundation model \citep{yimindou2025geological}.

	Practical deployment motivated lightweight fault networks and systematic refinements of CNN baselines. FaultSeg3D+ evaluated and improved CNN-based fault segmentation to address generalisation limits of the original synthetic-data paradigm \citep{li2024faultseg}. Other lightweight 3D designs introduced bidirectional decoding, dynamic scalability, or attribute-fusion modules to balance accuracy and efficiency \citep{tang2024fault,li2024faultb,yang2024multiurnet}. Transformer-based models continued to expand, including Transformer-enhanced fault detectors and volumetric Transformers for 3D horizon picking and horizon tracking \citep{zhou2024fault,liao2025automatic,zhao2025seismica}. Barely supervised strategies further pushed label efficiency through fault-orthogonal annotation and related schemes \citep{zhang2025seismic}.

	Horizon interpretation advanced via deep-learning methods with uncertainty encoding and vertical constraints, as well as production-oriented semi-automatic workflows with explicit uncertainty quantification \citep{liao2024deep,jung2025machine}. Complementary directions included geologically consistent stratal-surface construction guided by geologic time surfaces and probabilistic fault interpretation based on marked point processes \citep{wang2024stratal,tatymoukati2025fault}.

	Overall, the period reflects a transition from survey-specific models toward pretrainable, promptable, and cross-survey transferable frameworks. Self-supervised pretraining improves robustness to style and noise variation, SAM-style prompting enables human-in-the-loop editing, and lightweight plus label-efficient designs address production constraints. Looking further ahead, early efforts have begun to embed such interactive models within LLM-orchestrated agent workflows that translate natural-language goals into concrete interpretation operations \citep{kanfar2025intelligent,ren2025seismology}, although this direction is still highly exploratory.

	\begin{figure*}[!htb]
		\includegraphics[scale=0.51]{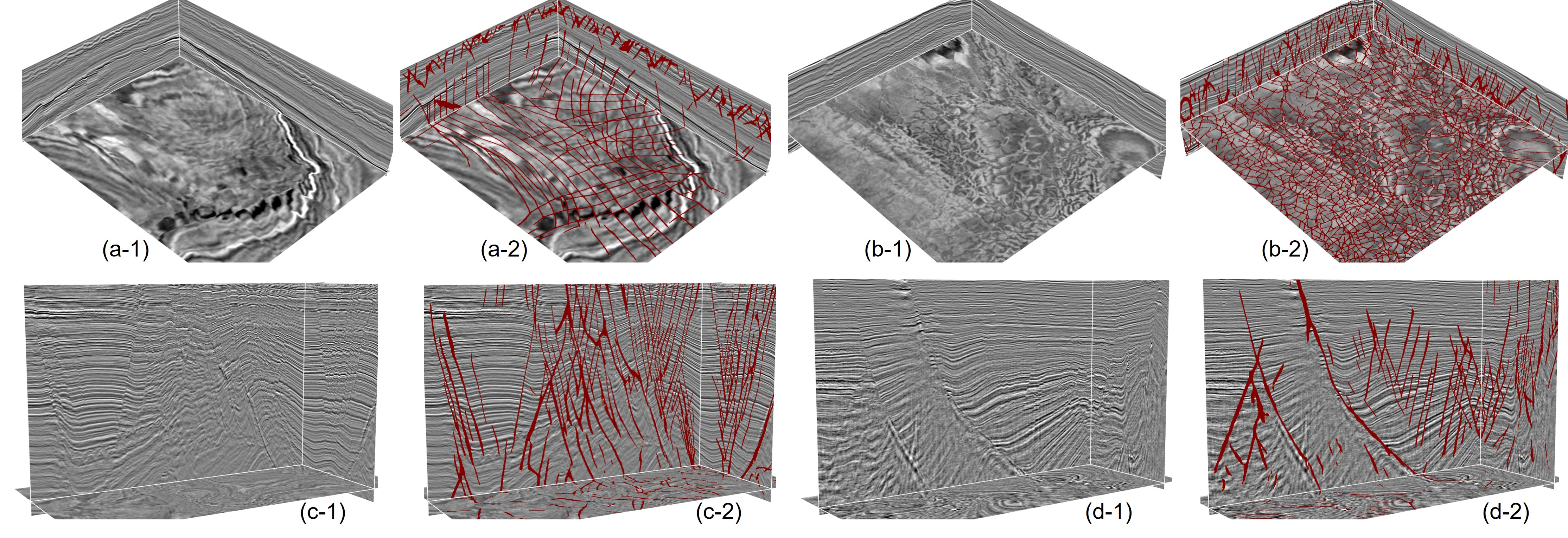}
		\centering\caption{Field outputs of the accompanying CIG-Bench fault baseline. The model is an HRNet backbone \citep{wang2021hrnet} with skip connections, trained on  synthetic dataset following the established FaultSeg3D synthetic-data training strategy \citep{wu2019faultseg}. The above raw seismic data are sourced from Nlog and the USGS.}
		\label{fig6}
	\end{figure*}

	\subsubsection{Summary and task-specific outlook}

	From an overall perspective, the development of seismic structural interpretation has largely followed a trajectory similar to that of computer vision, evolving from workflows based on manually designed seismic attributes and other handcrafted features toward end-to-end learning based interpretation \citep{bergen2019machine,yu2021deep,wrona2021seismic}. Unlike computer vision, however, the transition toward a relatively mature stage in structural interpretation has not been marked by any single network architecture. Instead, it has been enabled primarily by the proposal and widespread adoption of systematic synthetic data methodologies, which provide scalable labels, controlled structural variability, and reproducible evaluation protocols \citep{wu2019faultseg,cunha2020seismic}.

	Among the major task categories, fault related interpretation is currently the most widely deployed in practice, yet it still exhibits persistent limitations \citep{an2023current,li2024faultseg}. Generalization remains inadequate for deep intervals with weak seismic responses and low signal-to-noise ratios, as well as for complex tectonic styles such as strike slip systems with branching geometries and thrust related deformation. Under strong noise and acquisition footprints, automated models can also generate numerous false positives and spatially incoherent artifacts, which limits reliability in industrial settings.

	In comparison, horizon interpretation has not yet formed an equally universal solution despite progress in synthetic data and learning based pipelines \citep{tschannen2020extractinga,wu2019semiautomated}. Sparse marker horizon picking is often highly survey-specific and typically succeeds only for a small number of horizons with stable seismic responses, making it difficult to generalize cross-survey areas, stratigraphic sequences, and complex structural conditions \citep{peters2019multiresolution,lou2020seismic}. Methods with stronger transferability frequently rely on dense stratigraphic representations such as RGT to support correlation and horizon tracking \citep{bi2021deep,geng2020deep,di2022relative}. Nevertheless, accuracy and stability can degrade sharply in strongly deformed regions where strata are intensely disrupted and where unconformities and thrust nappes are prevalent, leading to reduced global consistency, axis crossing artifacts, and limited robustness under cross-survey distribution shifts \citep{geng2020deep,bi2021deep}. As a consequence, current approaches often struggle to meet industrial requirements for consistency and accuracy in complex structural settings.

	Looking ahead, fault interpretation research must continue to address several particularly challenging scenarios, including deep faults with weak seismic responses, strike slip faults and their branching systems, and strongly nonlinear deformation associated with thrusting and nappe emplacement \citep{an2023current,li2024faultseg}. Semi-automatic and interactive promptable paradigms can improve controllability and interpretation consistency through user guided constraints \citep{gao2026foundation}. However, for fault systems dominated by dense populations of small scale faults and complex spatial connectivity, heavy interaction can substantially increase manual workload and reduce overall efficiency. A promising direction is therefore to combine the throughput of fully automated segmentation with lightweight interactive refinement, where automation provides reliable initial fault volumes and targeted user inputs support rapid local correction and uncertainty guided editing.

	For horizon interpretation, RGT estimation is likely to remain the most general pathway for constructing 3D stratigraphic frameworks, provided that stronger constraints and consistency control are incorporated \citep{geng2020deep,bi2021deep}. Integrating geological and geophysical priors into learning, including fault geometry and displacement constraints, stratigraphic topological continuity, local monotonicity and smoothness, and anchoring to well tops or key interfaces, can suppress unrealistic oscillation and long range drift. In addition, sparse interactive inputs can serve as high value anchors to guide corrections in high uncertainty regions while preserving automatic efficiency elsewhere. A fused paradigm, with dense RGT as the backbone, physics consistent constraints as regulators, and sparse human anchors as guidance, is expected to improve global consistency and cross-survey generalization, and to provide a more stable representation for automatic horizon generation and subsequent structural modeling.

	\subsubsection{Illustrative example: fault and RGT field results}

\begin{figure*}[htb]
		\includegraphics[scale=0.45]{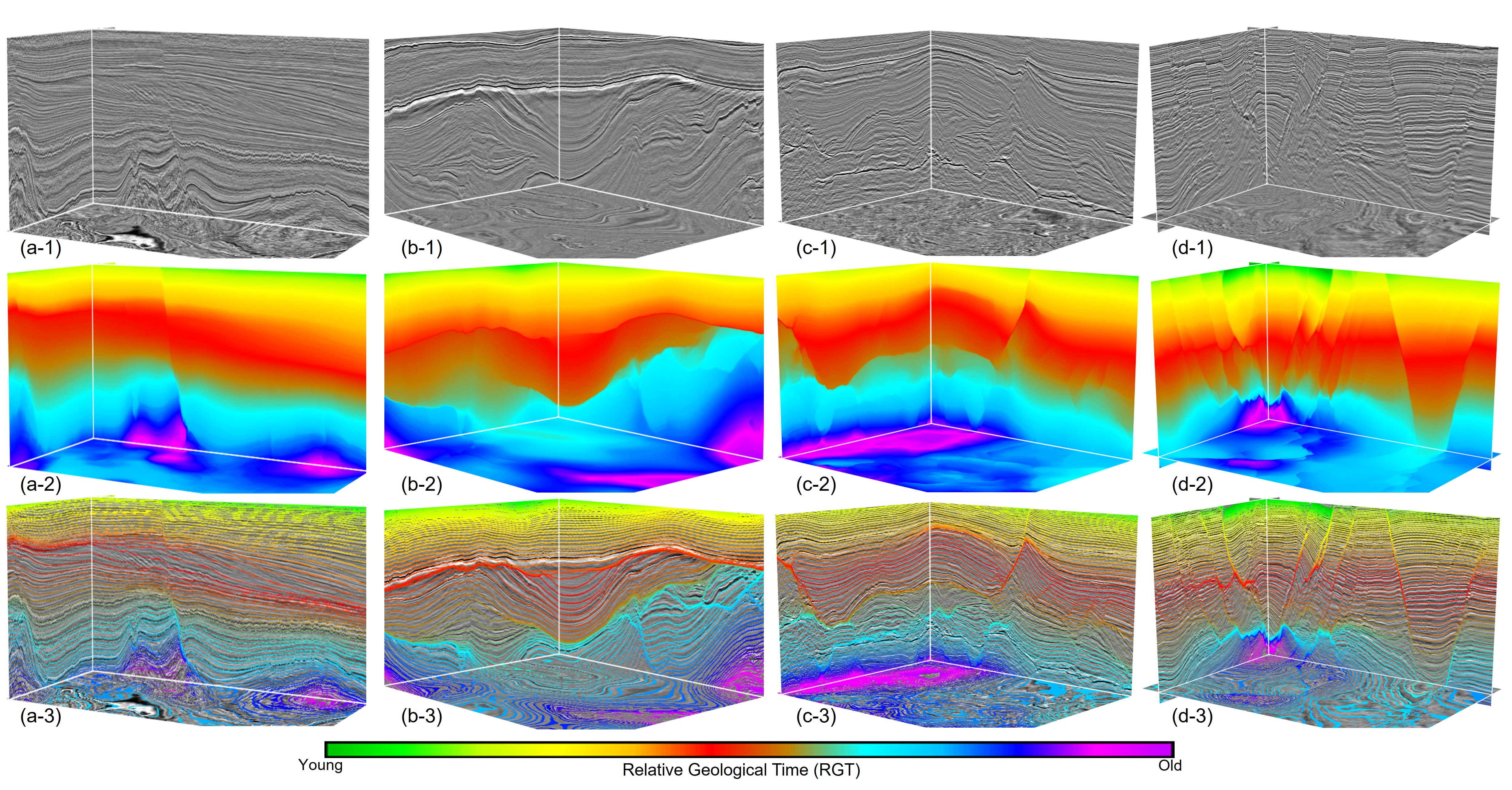}
		\centering\caption{Field outputs of the accompanying CIG-Bench RGT baseline, an HRNet backbone trained on synthetic dataset using the existing RGT-Est training strategy \citep{dou2026learning}; raw data from Nlog and the USGS. Top: seismic data. Middle: predicted RGT volume. Bottom: horizons extracted from the RGT volume, overlaid on the seismic data. Shown to illustrate that dense RGT yields stratigraphically consistent frameworks across complex settings.}
		\label{fig6-1}
	\end{figure*}

	To make the preceding discussion concrete, Figures~\ref{fig6} and~\ref{fig6-1}
	show representative field outputs of the accompanying CIG-Bench fault and RGT
	baselines; implementation details, inference APIs, and the corresponding
	synthetic quantitative comparison are given in the Supplementary Material
	(Sections~\ref{si:api}--\ref{si:quant}). These examples are intended to illustrate current
	capabilities and remaining limitations, not to establish state-of-the-art
	performance.

	Figure~\ref{fig6} shows fault segmentation on representative field volumes. On a fault-intensive area from the Netherlands F3 dataset, a classic fault-detection benchmark, the baseline delineates the major fault framework and also responds to small-scale polygonal faults in the shallow section. A second Netherlands example with intensive polygonal faulting yields a dense, spatially coherent polygonal network, capturing short, discontinuous, and intersecting fault segments. On USGS data with extremely dense faults and complex intersections, individual fault traces remain largely continuous, and for a deep-rooted fault system the dominant deep fault and associated secondary faults are recovered along the dipping plane. These examples suggest reasonable behaviour across both subtle polygonal faults and large-scale deep faults, while field deployment still faces the cross-survey generalization limits discussed throughout this section.

	Figure~\ref{fig6-1} shows RGT estimation on field data from Nlog and the USGS, using the RGT-Est training strategy \citep{dou2026learning}. The examples span slope bodies, unconformities, densely faulted zones, and multiple stratigraphic packages. The predicted RGT volumes remain smooth and stratigraphically consistent while preserving the main discontinuities caused by faults and erosional surfaces, and the extracted horizons follow seismic reflections with good lateral continuity across folded strata, steeply dipping layers, fault-controlled deformation, and unconformity-bounded sequences---illustrating that dense RGT remains a practical route to a 3D stratigraphic framework, consistent with the review above.

	\subsection{Geobody: Salt, Channel, and Karst Delineation}

	\begin{figure*}[htb]
		\includegraphics[scale=0.78]{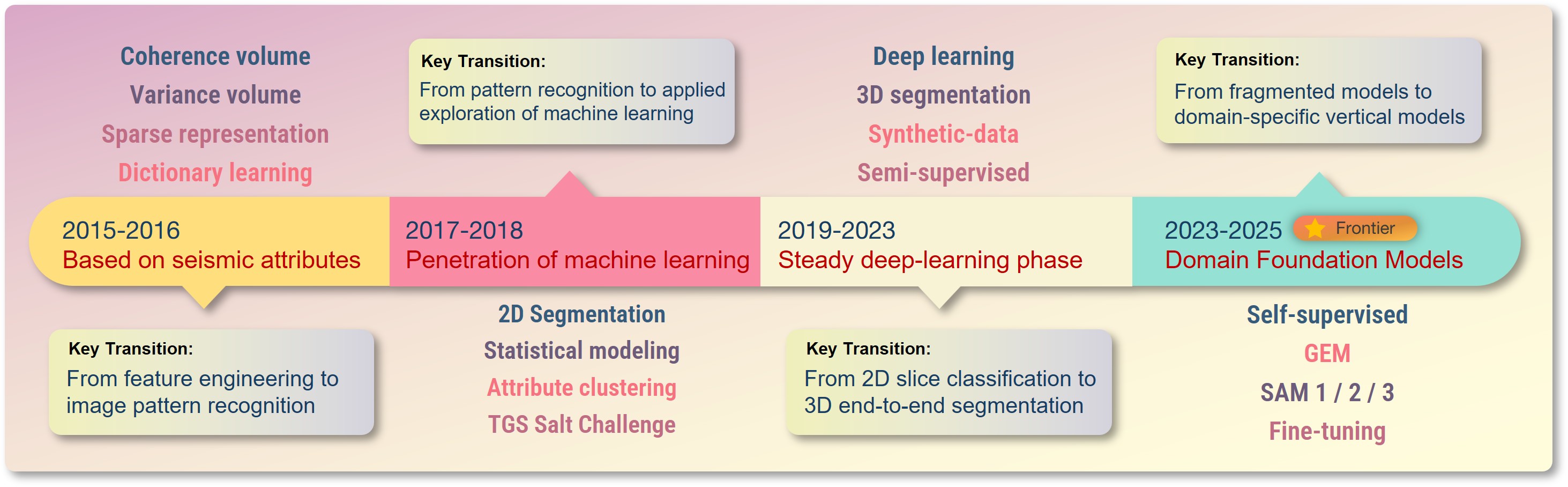}
		\centering\caption{The figure summarizes the technological evolution and paradigm shifts in geobody interpretation over the past decade, driven by advances in machine learning and deep learning. Using stage-specific keywords and representative milestones, it delineates the research focus of each phase and its central transitions.}
		\label{fig3}
	\end{figure*}

	Figure~\ref{fig3} summarizes the evolutionary trajectory of geobody interpretation over the past decade, from multiscale attribute enhancement and sparse-representation tools, through the early penetration of machine learning (accelerated by public competitions such as the TGS Salt Challenge), to the dominance of 3D end-to-end deep learning and, most recently, the exploration of domain foundation models with self-supervised pretraining and prompt-based interaction. Each phase is discussed in detail in the following subsections. Whether the foundation-model route proves durable remains to be confirmed through longer-term practical validation.

	\subsubsection{2015--2016: Seismic-attribute-driven interpretation}

	During this period, geobody identification remained grounded in attribute engineering as its primary technical foundation, yet it began to transition toward probabilistic-field modeling, supervised learning, and sparse representation frameworks \citep{wang2015noise,wu2016methods}. Early studies relied on multiscale and multidirectional seismic attributes—such as amplitude, envelope, spectral decomposition, and curvature—using attribute blending and visualization techniques to enhance the contrast between target geobodies and the surrounding background \citep{chinwuko2015coblending}. The central goal at this stage was to construct a three-dimensional interpretive workspace that aided human interpreters, rather than to achieve fully automatic geobody segmentation.

	Building on this foundation, researchers gradually introduced more systematic image-processing and pattern-recognition approaches to improve boundary detection and three-dimensional tracking robustness \citep{qi2015segmentation}. Representative strategies included applying edge-preserving filters to smooth multi-attribute volumes while maintaining geometric interfaces, followed by attribute-space clustering to separate targets from background. Other work exploited frequency-domain texture gradients, subspace learning, and noise-adjusted principal component analysis to ensure continuous boundary tracking even under strong noise and acquisition footprints \citep{wang2015noise}.

	At the same time, some studies reformulated geobody identification as a probabilistic-field estimation or optimization problem. One class of methods fused multiple attributes into a unified likelihood volume, from which geobody interfaces were extracted using three-dimensional image-analysis techniques \citep{wu2016methods}. Another class leveraged local texture features to train codebook or dictionary models, using sparse representation and supervised classification to achieve automatic segmentation on large-scale seismic volumes \citep{ramirez2016salt,amin2016salt}.

	Overall, this stage marked a shift from heuristic, threshold-based multi-attribute segmentation toward more optimizable and verifiable mathematical formulations, laying the conceptual groundwork for the emergence of deep-learning-based geobody interpretation in subsequent years.

	\subsubsection{2017--2018: Early adoption of machine learning}

	From 2017 to 2018, geobody identification entered an early machine-learning stage, as convolutional neural networks were first applied to salt classification while attribute-based and clustering workflows continued to develop \citep{waldeland2018convolutional,waldeland2017salt}.

	For salt bodies, attribute clustering remained common. \citet{di2018multi} proposed a multi-attribute $k$-means workflow that computes amplitude, phase, texture, and related attributes and then performs unsupervised partitioning for salt-boundary delineation. The key shift was the adoption of CNNs: \citet{waldeland2017salt} showed that a CNN can unify feature extraction and classification for salt, with training on a single labelled inline slice already sufficient to classify a full volume, outperforming attribute-then-classify pipelines, and \citet{waldeland2018convolutional} generalised this idea toward automated seismic interpretation by taking raw seismic amplitudes directly as input so that the network learns its own attributes rather than relying on hand-engineered ones, becoming the most cited geobody work of the period and establishing CNNs as a viable paradigm for geobody segmentation. Public initiatives such as the TGS Salt Identification Challenge on Kaggle further accelerated community uptake of 2D segmentation networks for salt interpretation.

	For channel geobodies, research still emphasised multiscale and multidirectional attribute characterisation. Shearlet transforms were used to detect channel boundaries by exploiting anisotropy across scales and orientations \citep{karbalaali2017channela,karbalaali2018seismic}, while GLCM-based texture attributes and directional structure-tensor coherence provided complementary cues for channel and discontinuity detection \citep{mohebian2018detection,wu2017directional}. Deep learning had not yet become central for channel interpretation in this period.

	Karst geobodies saw early machine-learning explorations. Statistical neural-network imaging of karst systems in 3D seismic addressed complex void geometries \citep{ebuna2018statistical}, and an optimised CNN was proposed for karst-cave reservoir identification \citep{cai2018identification}. Overall, 2017--2018 marked a transition from feature engineering to image-pattern recognition, most prominently for salt, while channels largely remained attribute-driven and karst began to adopt early neural approaches.

	\subsubsection{2019--2023: Deep-learning consolidation}

	From 2019 to 2023, geobody identification entered a relatively mature stage dominated by deep learning, with geobody extraction increasingly formulated as end-to-end 3D segmentation using encoder--decoder networks for voxel-level prediction \citep{pham2019automatic,gao2021channelseg}. Salt-body segmentation advanced most rapidly. \citet{shi2019saltseg} introduced SaltSeg, which casts salt-body interpretation as 3D image segmentation using an encoder--decoder CNN trained on binary salt/non-salt labels, with a data generator that extracts randomly positioned subvolumes from large 3D training data and applies augmentation (the most cited work of this period). Subsequent studies explored production-scale pipelines, improved UNet variants for joint salt and fault detection, and interactive salt segmentation, alongside systematic comparisons of architectures and regularisation strategies for stable training and better generalisation \citep{sen2020saltnet,alfarhan2022robust,di2020comparison}. Unlike faults, however, purely synthetic-data training was less straightforward for salt due to semantic ambiguity in amplitudes and high morphological diversity \citep{warren2023toward}.

	Label scarcity motivated extensive semi-supervised and weakly supervised learning for salt interpretation. Ensemble- and teacher-student-based frameworks, iterative pseudo-labelling, boundary-aware losses, and edge-guided branches improved performance under sparse or coarse labels \citep{babakhin2019semi,geng2022semisupervised}. Knowledge distillation variants were also explored to compress and enhance salt models, and semi-supervised segmentation was extended to broader seismic interpretation tasks \citep{li2023salt,wang2023semisupervised}.

	Channel interpretation similarly transitioned to deep learning. \citet{pham2019automatic} demonstrated automatic 3D channel detection using an encoder--decoder (SegNet) network, with a Bayesian SegNet variant providing uncertainty estimates; trained purely on synthetic volumes, it transferred to field data from the Browse Basin (offshore Australia) and Parihaka (New Zealand). ChannelSeg3D addressed the absence of labelled field channels by generating training data through process-based channel simulation, paired with 3D segmentation networks \citep{gao2021channelseg,gao2020channel}. ResNet-style methods were applied to complex channel settings, while uncertainty and interpretability analyses addressed reliability requirements for automated workflows \citep{li2022resnet,pham2021uncertainty}. Attribute-combination strategies and ANN-based approaches remained complementary for specific channel-related targets \citep{khasrajinejad2021proposing,ismail2022channels}.

	Karst interpretation emerged as a smaller but growing sub-field. Bayesian deep learning provided probabilistic characterisation of paleocaves, tree-based models were used for cavity identification from impedance images, and CNNs with transfer learning targeted the scarcity of labelled karst data \citep{zhang2022seismic,kouassi2023identification,huang2023automatic}. Self-supervised and interactive approaches also appeared, including label-free structure delineation via latent-space projection and flood-filling networks for expert-in-the-loop geobody tracking \citep{aribido2021self,shi2021interactively}.

	Overall, 2019--2023 established 3D end-to-end segmentation as the dominant paradigm, supported by advances in architectures, losses, and label-efficient learning.

	\subsubsection{2024--2025: Emergence of domain foundation models}

	From 2024 to 2025, geobody interpretation was increasingly re-oriented around domain foundation models, which aim to learn unified, transferable subsurface representations that support geobody segmentation cross-surveys and geobody types within a single pretrained backbone \citep{sheng2025seismic,gao2026foundation,islam2024comprehensive}. Rather than advancing primarily through task-specific network tweaks, the field shifted toward scalable pretraining on large seismic corpora, prompt-driven interaction, and standardised evaluation resources that together enable stronger cross-survey generalisation and more practical deployment.

	In this context, the seismic foundation model pretrained on 192 global surveys demonstrated that large-scale pretraining can deliver robust transfer to downstream interpretation tasks, including geobody segmentation, under limited labelled data \citep{sheng2025seismic}. Complementing this, foundation architectures equipped with multi-modal prompt engines extended segmentation from passive automation to expert-in-the-loop workflows, allowing different prompt modalities to control and refine predictions across diverse geobody categories \citep{gao2026foundation}. Cross-domain adaptation from computer-vision foundation models to geophysical imagery further reinforced the premise that reusable pretrained representations can mitigate chronic data curation constraints in geoscience \citep{guo2025cross}.

	Recent unified promptable frameworks have synthesised these ideas to span structural interpretation, geobody delineation, and property interpretation within a single model, illustrating a broader move toward general-purpose subsurface understanding rather than single-task pipelines \citep{yimindou2025geological}. Alongside this model shift, the release of large-scale benchmark datasets provided the infrastructure needed to train and compare foundation-model adaptations reproducibly, while Transformer-based and interactive segmentation methods were increasingly integrated as modular components within foundation-model-centric workflows \citep{wang2025cigchannel}.

	Overall, a salient feature of this period is the growing exploration of geobody interpretation around domain foundation models, where pretraining, prompting, and benchmarking jointly aim at the cross-survey scalability that earlier deep-learning approaches struggled to achieve.

	\subsubsection{Summary and task-specific outlook}\label{sec:geobody_outlook}

	Compared with the other three interpretation tasks, geobody analysis has received relatively less attention \citep{islam2024comprehensive}. In traditional workflows, geobodies are often treated as secondary products derived after inversion and attribute modeling via thresholding, clustering, or connectivity analysis. If geobodies are treated as explicit targets, large-scale manual annotation is usually required, and both label acquisition and quality control are difficult \citep{warren2023toward}. Although deep learning segmentation enables direct identification of salt bodies, channels, and karst systems \citep{shi2019saltseg,pham2019automatic,yan2025karst}, cross-survey generalization remains challenging due to heterogeneous imaging quality, differences in acquisition and processing, and highly variable morphology \citep{di2020comparison,sen2020saltnet}. These factors have hindered the formation of mature benchmarks and stable engineering workflows comparable to fault interpretation \citep{wu2019faultseg}.

	Accordingly, engineering level deployment of AI based geobody interpretation remains limited, mainly due to weak cross-survey robustness and limited adaptability to complex morphologies \citep{geng2022semisupervised,babakhin2019semi}. Interactive promptable paradigms can improve controllability by using sparse prompts to guide rapid refinement cross-surveys and geobody types, partially mitigating the instability of fully automatic methods \citep{gao2026foundation}. Nevertheless, robust generalization ultimately requires broader and more realistic 3D datasets with standardized protocols, together with richer geological and geophysical priors \citep{sheng2025seismic,wang2025cigchannel}.

	Looking ahead, a priority is to establish a comprehensive 3D benchmark suite that covers production level scale, resolution, and signal-to-noise ratio, and is representative across geobody types, morphological complexity, structural context, and imaging conditions, supported by unified splits, guidelines, and metrics. Methodologically, because geobody boundaries are intrinsically ambiguous under bandwidth limits and noise, future models should incorporate priors such as shape and connectivity constraints, stratigraphic and structural controls, attribute consistency regularization, and sparse anchoring to well logs or existing interpretation products. A practical pathway is to combine automatic segmentation with lightweight interactive refinement, enabling efficient initial interpretations and targeted correction in high uncertainty regions, and ultimately producing stable, transferable models that support downstream tasks including structural modeling and reservoir characterization.

	\subsubsection{Illustrative example: geobody field results}\label{sec:cigbench_geobody}

	The accompanying CIG-Bench resource currently provides geobody baselines for channels and karst caves; salt bodies, although extensively discussed in the Geobody review above, are not included in the present release, because purely synthetic-data training has proven less straightforward for salt due to semantic ambiguity in amplitudes and high morphological diversity \citep{warren2023toward}, and a unified synthetic salt-generation pipeline consistent with the channel and karst pipelines is still under development. Implementation details and inference APIs are given in the Supplementary Material (Section~\ref{si:api}).

\begin{figure*}[htb]
		\includegraphics[scale=0.55]{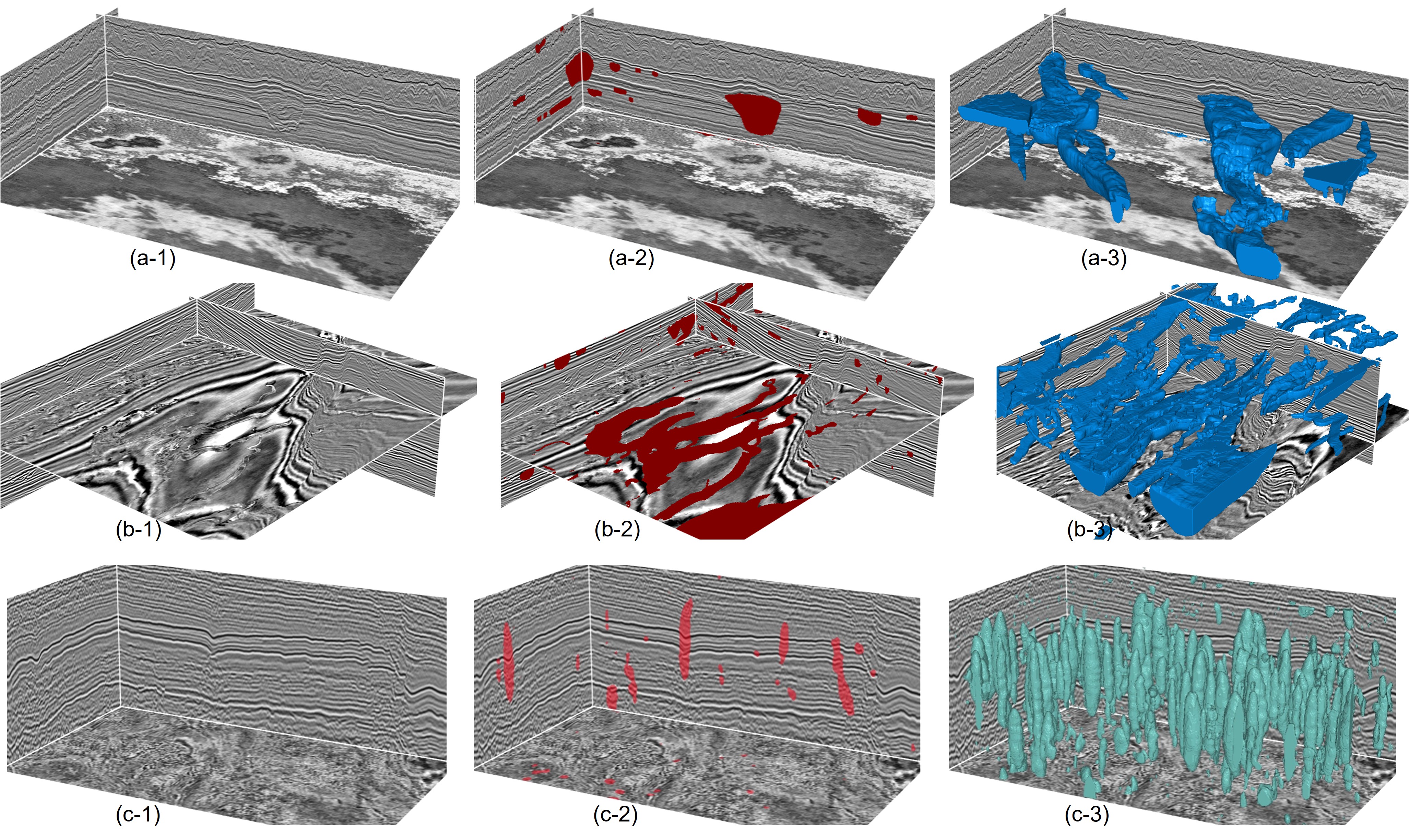}
		\centering\caption{Field outputs of the accompanying CIG-Bench channel and karst baselines. Both models use an HRNet backbone  trained on synthetic dataset, following the established ChannelSeg \citep{gao2021channelseg} and KarstSeg \citep{wu2020deepkast} synthetic-data training strategies, respectively. Panels (a), (b), and (c) show examples from the New Zealand Romney, the New Zealand Parihaka, and the USGS G3D, respectively.}
		\label{fig7}
	\end{figure*}

	Figure~\ref{fig7} shows field outputs of the channel and karst baselines on the New Zealand Parihaka, New Zealand Romney, and USGS G3D datasets (panels a--c); in each row the left column is the seismic volume, the middle the predicted segmentation overlaid on seismic data, and the right the extracted 3D geobodies.

	These examples are most useful for what they reveal about the open problems emphasised in the Geobody review. The baselines recover several channel- and karst-related geobodies and give a preliminary 3D picture of their spatial distribution, but fully automatic channel detection remains hard, especially across surveys: channel bodies show weak responses, discontinuous boundaries, strong lateral variability, and amplitudes similar to surrounding features. Accordingly, the channel results still contain missed detections, fragmented bodies, and false positives, confirming that robust cross-survey channel segmentation is unsolved. The karst examples (panel c) show more distinct vertical, clustered responses but still exhibit over-segmentation and incomplete boundaries---concretely illustrating why standardized 3D geobody benchmarks and stronger morphological priors are needed.

	\subsection{Facies: Depositional and Lithofacies Interpretation}
	\begin{figure*}[htb]
		\includegraphics[scale=0.78]{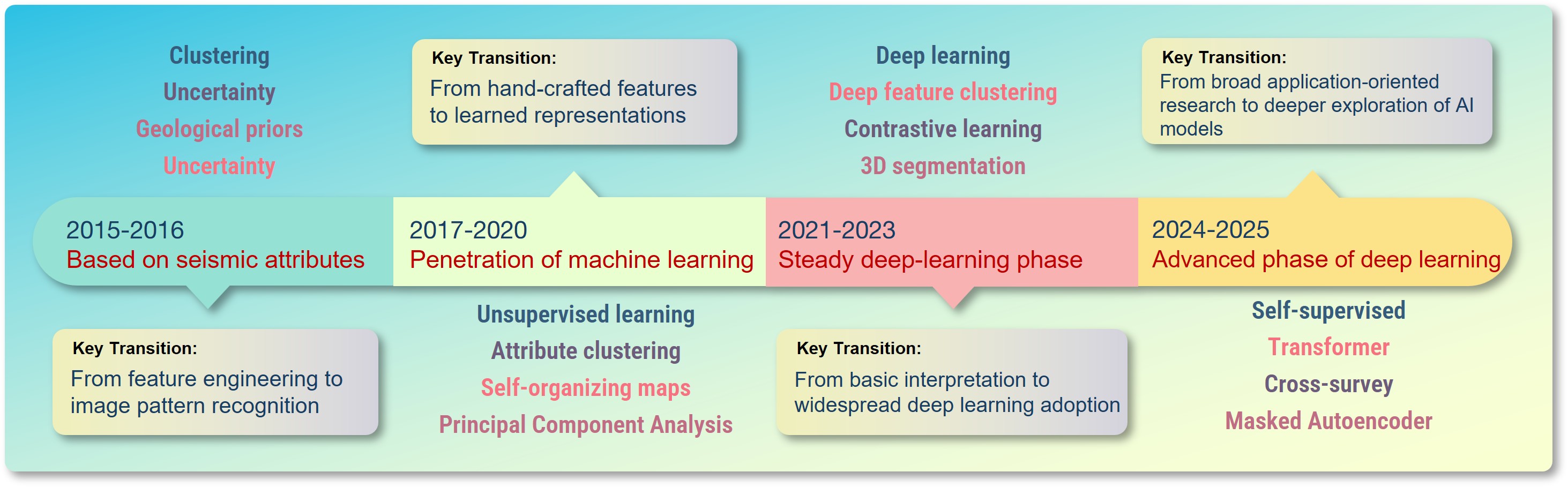}
		\centering\caption{The figure summarizes the technological evolution and paradigm shifts in facies interpretation over the past decade, driven by advances in machine learning and deep learning. Using stage-specific keywords and representative milestones, it delineates the research focus of each phase and its central transitions.}
		\label{fig4}
	\end{figure*}

	Figure~\ref{fig4} summarizes the evolution of seismic facies interpretation over the past decade, from seismic-attribute clustering and probabilistic characterization, through the adoption of unsupervised multi-attribute learning, to deep feature clustering, contrastive learning, and 3D segmentation, and most recently toward self-supervised learning, Transformers, and pretraining paradigms aimed at more generalizable and reusable representations. The individual phases are detailed in the following subsections.

	\subsubsection{2015--2016: Seismic-attribute-driven interpretation}

	From 2015 to 2016, seismic-facies research remained in a seismic-attribute--based stage, where the central objective was to design discriminative attribute combinations and classify facies in multi-attribute space. Coherence, variance, amplitude, phase, and texture attributes were widely computed, and multiscale, multidirectional responses---such as Kuwahara-filtered volumes---were used to sharpen contrasts between neighbouring facies groups, making attribute design, combination, and visualization the determining factors for methodological performance \citep{qi2016semisupervised}. Beyond absolute amplitude magnitudes, amplitude variation patterns were also explored as classification inputs, improving the identification of facies in weak-amplitude areas \citep{liu2015facies}.

	Algorithmically, multi-attribute values from voxels or small windows were concatenated into feature vectors, and clustering was performed in attribute space using k-means, self-organising maps (SOM), generative topographic mapping (GTM), support vector machines (SVM), Gaussian mixture models, and artificial neural networks \citep{zhao2015attribute}. \citet{zhao2015comparison} systematically compared six such classifiers on a single turbidite data set and concluded that, although supervised methods yielded accurate estimates of desired facies, unsupervised methods could also highlight features that might otherwise be overlooked. Building on this, \citet{qi2016semisupervised} proposed a semisupervised workflow in which the interpreter manually painted target facies to generate training data; candidate attributes were evaluated by cross-correlating their per-facies histograms, and Kuwahara filtering significantly increased discrimination. Unlabelled voxels were then classified via probability density functions projected on a GTM manifold, producing a probability volume for each user-defined facies.

	To maintain geological plausibility, several works explicitly incorporated geological priors into the classification or post-processing stage. \citet{gonzalez2016adding} demonstrated that, in a turbidite setting with sparse well control, augmenting the initial lithofluid-facies model with rock types expected from the depositional environment---but not sampled at the well---and combining spatially variant prior probabilities with data likelihoods through Bayesian estimation substantially reduced prediction bias. Probabilistic facies volumes and entropy-based uncertainty metrics were further introduced to shift facies interpretation away from single hard classifications toward quantitative probabilistic descriptions \citep{yuan2016quantitative}.

	Notably, a first attempt at applying deep learning to seismic-facies recognition also appeared during this period: \citet{li2016recognition} used a Deep Belief Network (DBN) stacked from Restricted Boltzmann Machines to extract features directly from multi-sample seismic inputs for lithology recognition, foreshadowing the data-driven feature-learning paradigm that would dominate later years.

	Overall, the feature representations at this stage were still predominantly hand-crafted, and most seismic-facies problems had been reformulated as clustering and probabilistic-modelling tasks in high-dimensional attribute space. The coexistence of traditional machine-learning classifiers and the nascent introduction of deep architectures such as DBN laid the methodological groundwork for the subsequent rapid adoption of deep-learning techniques.

	\subsubsection{2017--2020: Early adoption of machine learning}
	From 2017 to 2020, seismic-facies analysis entered an early-adoption stage of machine learning, during which data-driven approaches expanded from traditional clustering to deep-learning architectures \citep{ross2017comparison,alaudah2019machine}. For example, \citet{wrona2018seismic} built a rigorous, repeatable attribute-based machine-learning workflow on 3D broadband data from the northern North Sea, extracting seismic attributes and training a classifier from a manually labelled subset to propagate facies through the volume. Methods in this period generally followed two perspectives: a stratal (attribute-driven) perspective and a profile-based (image-segmentation) perspective. The stratal perspective interprets facies along horizon-aligned or stratigraphically constrained slices, which better conforms to depositional architecture and can yield more geologically consistent delineation \citep{zhao2017constraining,zhao2018seismic}. However, it typically requires heavier data preparation, stronger structural constraints, and greater interpretive effort. In contrast, the profile-based perspective performs facies partitioning on inline or crossline sections and formulates the problem as image segmentation \citep{liu2020seismic,zhang2020seismica}. This workflow is more streamlined and computationally efficient, but it often produces coarser results and has limited ability to resolve fine facies boundaries and internal variability.

	Under the stratal perspective, many studies combined multi-attribute feature vectors with clustering or supervised classification. A typical workflow computes diverse seismic attributes, including texture descriptors such as the gray-level co-occurrence matrix (GLCM) \citep{di2017nonlinear}, and then clusters or classifies voxels using k-means, self-organising maps, or random forests \citep{kim2018seismic,li2019fusing}. Multi-waveform classification was also proposed to improve stability and stratigraphic continuity \citep{song2017multi}, and alternative feature-extraction strategies, including speech-recognition-inspired parameters, were introduced to enhance sensitivity to reservoir characteristics \citep{ebuna2018statistical}. Several works emphasised geological constraints by reassigning or smoothing labels using facies proportions, sequence boundaries, or palaeoenvironmental priors, and by incorporating spatial regularisation (e.g., Markov random fields or geostatistical filtering) to improve continuity \citep{zhao2017constraining,qi2020seismic}. Attribute selection received increasing attention, with data-adaptive weighting and SHAP-based interpretability analyses used to identify discriminative subsets \citep{zhao2018seismic,luborobles2020machine}.

	In parallel, image-segmentation--based facies analysis rose rapidly, using inline or crossline profiles as primary inputs \citep{alaudah2019machine,ross2017comparison}. Public benchmark releases, particularly the Netherlands F3 block, reduced the barrier to supervised learning and accelerated method development; \citet{alaudah2019machine} explicitly targeted the absence of large, publicly available annotated facies datasets and the resulting inconsistency in classes, train/test splits, and quantitative reporting by releasing a standardized machine-learning benchmark. In particular, \citet{zhao2018seismicCNN} was among the first to introduce convolutional neural networks into seismic-facies classification, systematically comparing several CNN architectures and demonstrating their effectiveness for profile-based facies partitioning. Many studies adopted computer-vision architectures, including CNNs, encoder--decoder networks, and GANs, to enable efficient facies partitioning and volume-wide propagation \citep{liu2020seismic,zhang2019automatic,grana2020comparison}. Deep convolutional autoencoders further enabled unsupervised facies analysis on prestack data via end-to-end feature learning without manually designed attributes \citep{qian2018unsupervised,duan2019seismic}. Semi-supervised GAN-based strategies mitigated label scarcity \citep{liu2020seismic}, while Bayesian deep-learning approaches using Monte Carlo dropout provided voxel-wise uncertainty estimates alongside predictions \citep{mukhopadhyay2019bayesian}. Comparative studies suggested that deep learning and Monte Carlo--based probabilistic methods can reach similar accuracy, while offering complementary advantages in uncertainty characterisation \citep{grana2020comparison}.

	Overall, this period marks the first rapid expansion of machine learning in seismic-facies interpretation \citep{luborobles2019independent,feng2020lithofacies}. Compared with structural targets such as faults or horizons, facies are inherently more ambiguous and variable, depending on depositional environment, scale, and interpretive objectives. Accordingly, studies adopted divergent definitions of ``facies'' and heterogeneous technical pathways, underscoring the need for standardised benchmarks and reproducible evaluation protocols, a gap that the F3 dataset began to address \citep{alaudah2019machine}.

	\subsubsection{2021--2023: Deep-learning consolidation}

	From 2021 to 2023, seismic-facies research entered a relatively mature stage dominated by deep learning \citep{li2021addcnn,feng2021bayesian,zhang2021automatic}. The two methodological perspectives established earlier---profile-based segmentation and stratal-perspective analysis---continued to advance in parallel, while label-efficient learning became a central theme.

	Driven by the convenient organisation of inline/crossline profiles, the availability of open benchmark datasets (e.g.\ SEG 2020 open data), and natural compatibility with semantic segmentation, profile-based facies segmentation grew rapidly \citep{xu2022seismic,chai2022open}. Architectures evolved from UNet variants \citep{tolstaya2022deep} to attention-augmented CNNs \citep{li2021addcnn}, DeepLabv3+/GAN-based frameworks \citep{kaur2023deep}, and vision transformers such as SegFormer-style encoder--decoder designs \citep{wang2023seismicb}. Beyond architecture, tailored losses and training strategies were introduced to address class imbalance, improve detection of thin facies and subtle boundaries, and preserve geometric continuity \citep{chen2022stronger,zhan2023subsurface}. Explainability also received increasing attention, with attention maps and intrinsic interpretation methods used to identify seismic cues underlying model decisions and to mitigate the ``black-box'' concern \citep{noh2023explainable,li2021addcnn,luborobles2022quantifying}.

	For stratal-perspective 3D facies analysis, research largely transitioned from hand-crafted attributes to deep feature representations, replacing attribute clustering with deep feature clustering \citep{puzyrev2022unsupervised,li2023unsupervised}. Deep features enabled joint representation of structural context and depositional variability, facilitating modelling of continuous facies belts and quasi-3D stratigraphic units. RGT was incorporated as an explicit constraint in CNN-based facies classification to enforce stratigraphic ordering and depositional continuity \citep{di2021relative}. Semi-supervised fuzzy clustering on extended elastic impedance was also explored to bridge well-scale and seismic-scale facies definitions \citep{mirzakhanian2022semisupervised}.

	A defining theme was reduced reliance on manual labels. Deep feature clustering and contrastive learning were widely adopted, often via unsupervised pretraining (e.g.\ autoencoding or masked reconstruction) followed by contrastive refinement across profiles or stratigraphic units \citep{li2023unsupervised,li2023conss}. Semi-supervised learning combined limited labels with large unlabelled volumes using autoencoders or GANs \citep{liu2021semi,xu2023deep}. Active learning was proposed to prioritise informative samples for annotation \citep{mustafa2023active}, while few-shot and prototype-based methods targeted rapid adaptation to new surveys \citep{zhao2023shot}. Domain adaptation further addressed cross-survey generalisation by transferring representations from labelled source data to unlabelled targets \citep{nasim2022seismic}.

	Bayesian deep-learning frameworks also matured, providing calibrated voxel-wise uncertainty estimates \citep{feng2021bayesian}, while classical probabilistic approaches such as Markov chain Monte Carlo remained relevant for posterior facies inference \citep{grana2023markov}. Overall, progress during this period was largely incremental---improving architectures, label efficiency, and uncertainty quantification---while persistent challenges, including cross-survey generalisation, label scarcity, and geological interpretability, were increasingly recognised but not yet systematically resolved \citep{xu2022seismic,babikir2022evaluation}.

	\subsubsection{2024--2025: Advanced deep-learning models}

	The 2024--2025 period is marked by foundation-model adaptation, automated architecture design, and multi-task learning, collectively pushing seismic-facies interpretation toward higher efficiency, stronger generalisation, and greater interactivity \citep{chikhaoui2024self,gao2024optimizing,guo2025seismic}.

	A key development is the transfer of large pretrained vision models to facies segmentation. FaciesSAM adapts the Fast Segment Anything Model by decomposing the workflow into 'Segment All' and prompt-guided 'Segment One', enabling human-in-the-loop interpretation with improved controllability and editability \citep{atolagbe2025toward}. More broadly, general-purpose models such as the seismic foundation model \citep{sheng2025seismic} and the Geological Everything Model 3D \citep{yimindou2025geological} demonstrate that representations pretrained on large seismic corpora can transfer to facies segmentation and other downstream tasks in zero-shot or few-shot regimes, suggesting a possible shift from task-specific training toward unified, promptable subsurface understanding---although the long-term viability of this direction, largely borrowed from the success of foundation models in other fields, still awaits broader practical validation.

	AutoML has also gained visibility. Neural architecture search (NAS) methods such as PC-DARTS-SFC use differentiable search to discover architectures better suited to facies classification than hand-crafted designs, outperforming the original PC-DARTS on the F3 benchmark and highlighting the practical value of automated optimisation in geophysical settings \citep{gao2024optimizing}. Meanwhile, incremental architectural refinements, including improved deep dilated CNNs with expanded receptive fields, continue to yield measurable gains in boundary delineation \citep{yang2024improved}.

	Self-supervised learning (SSL) matured into a practical response to label scarcity. \citet{chikhaoui2024self} show that reconstruction-based SSL pretraining followed by fine-tuning can reach performance comparable to fully supervised learning using only 5\%--10\% labelled data on the F3 and Penobscot datasets, while also improving cross-domain adaptation under limited annotation. In parallel, the release of cigFacies provides substantially larger and more diverse data resources for both supervised and self-supervised paradigms \citep{gao2025cigfacies}.

	Additional trends include spatiotemporal modelling and multi-task coupling. Convolutional LSTM-style designs capture spatial structure and inter-slice dependencies to improve delineation of thin interbeds and weak interfaces \citep{tian2025enhancing}. Multi-task frameworks such as SRSS-Net jointly perform super-resolution and segmentation, stabilising facies prediction under low-resolution inputs \citep{guo2025seismic}. RGT-constrained workflows continued to mature, with sensitivity analyses indicating improved accuracy and continuity and robustness to moderate RGT errors \citep{wu2024sensitivity}. Graph-based and hybrid methods further integrate geological adjacency and traditional constraints, while Bayesian and stochastic approaches remain relevant for uncertainty-aware prediction \citep{alswaidan2024geology,wang2024seismic,fernandes2024stochastic}.

	Although this period saw a series of advances, it did not resolve the fundamental challenges of seismic facies research in the machine learning setting, most notably how to define facies and how to establish standardized datasets and evaluation protocols.

	\subsubsection{Summary and task-specific outlook}

	Seismic facies link structural interpretation, geobody analysis, and property interpretation, and therefore occupy a central position in the interpretation workflow \citep{xu2022seismic,wrona2018seismic}. Although seismic facies has attracted far more research than geobodies, progress has mostly been incremental within established deep learning pipelines, largely emphasizing feature extraction and segmentation architectures rather than redefining the task or the interpretational paradigm \citep{qian2018unsupervised}.

	Most profile based studies still rely on a few public datasets and representative surveys such as F3 and Parihaka \citep{alaudah2019machine,chai2022open}, which makes models highly dataset dependent and weakly transferable across basins and data conditions \citep{li2021addcnn,liu2020seismic,zhang2021automatic}. A key bottleneck is the lack of an operationally consistent task definition: sectional and horizon based facies depend strongly on scale, label semantics, and geological context, so labeling systems and evaluation protocols vary widely and results are often not comparable or reproducible.

	We argue that the priority is to establish an evaluable, cross-survey task definition and benchmark framework \citep{sheng2025seismic,yimindou2025geological,gao2025cigfacies}. This requires reproducible annotation guidelines and hierarchical labels, representations that reconcile sectional, horizon based, and volumetric views, and metrics that emphasize cross-survey generalization, semantic consistency, boundary uncertainty, and interpretation utility rather than within dataset pixel level accuracy. Shared community standards in both methodology and evaluation are essential to move seismic facies analysis from experience driven practice toward a systematic and transferable paradigm.

	\subsubsection{Why no facies baseline is provided}
	Because seismic facies are conceptually ill-defined, their delineation often depends on interpreter expertise, research objectives, and the geological context of a given survey area, which makes it difficult to establish consistent class boundaries and annotation criteria. Moreover, existing studies differ substantially in the number of facies classes, labeling taxonomies, interpretation scale (sectional view versus horizon-based view), and supervision regime (fully supervised, weakly supervised, semi-supervised, or unsupervised). As a result, neither the methodological landscape nor the evaluation protocols have converged to a widely accepted standard.

	Under these conditions, the accompanying resource deliberately does not provide a single ``unified facies baseline'' directly reusable across datasets and labeling systems. Doing so would risk producing non-comparable performance conclusions under inconsistent label semantics, which would undermine the credibility and interpretability of any benchmark. We anticipate that seismic-facies research will move beyond survey- and interpreter-dependent empirical practices toward a standardized framework in which both data and methods are more clearly defined, reproducible, and comparable.

	\subsection{Property: Elastic and Petrophysical Property Inversion}
	\begin{figure*}[htb]
		\includegraphics[scale=0.78]{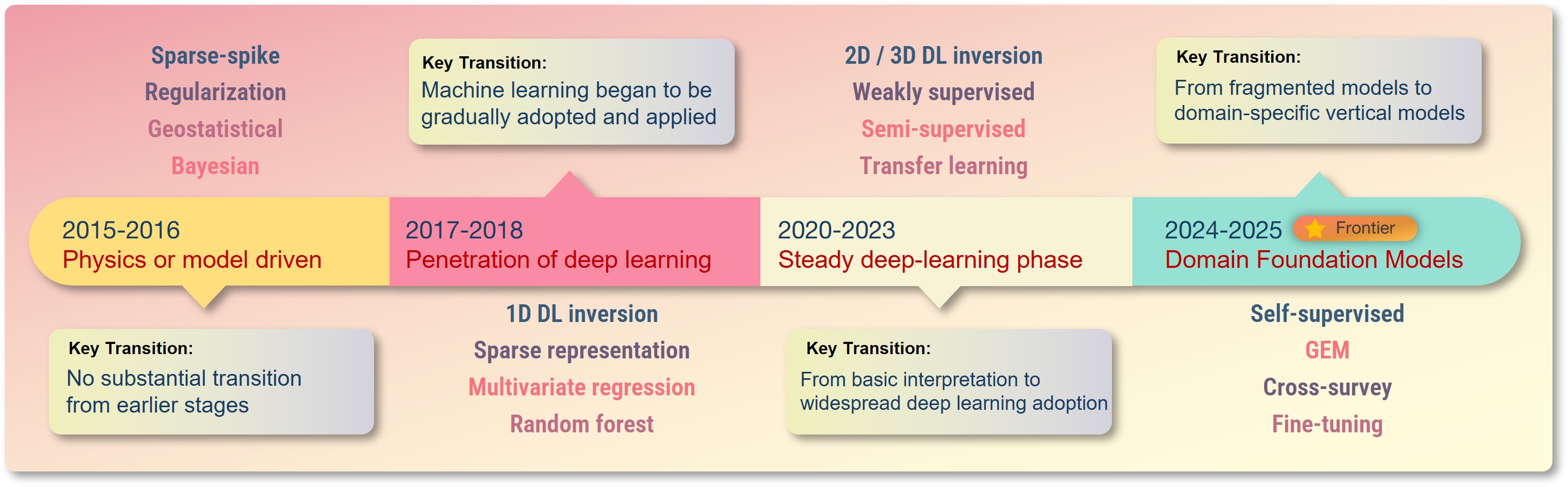}
		\centering\caption{The figure summarizes the technological evolution and paradigm shifts in property interpretation over the past decade, driven by advances in machine learning and deep learning. Using stage-specific keywords and representative milestones, it delineates the research focus of each phase and its central transitions.}
		\label{fig5}
	\end{figure*}

	Figure~\ref{fig5} summarizes the evolution of property interpretation over the past decade, from physics- and model-driven constrained inversion, through the emergence of learning-based 1D trace-level estimation, to 2D/3D end-to-end deep inversion with transfer and weak/semi-supervised learning, and most recently toward cross-survey generalization and prompt-driven, pretraining-based unified formulations \citep{yimindou2025geological}. Each phase is detailed in the following subsections; whether the most recent prompt-driven approaches can be sustained and reliably deployed in practice still requires longer-term validation.

	\subsubsection{2015--2017: Physics- and model-driven paradigm}

	During 2015--2017, seismic-property research largely remained within a physics- and model-driven paradigm, extending classical geophysical inversion with limited reliance on learning-based representations \citep{bui2017introduction,grana2016bayesian}. Most studies cast the inversion of acoustic impedance, velocity, porosity, or fluid-related properties as constrained optimisation problems grounded in the wave equation or AVO theory. Objective functions typically combined a data-misfit term with regularisation promoting sparsity, smoothness, or spatial continuity \citep{gholami2016fast,li2017seismic,kumar2016methodology}.

	A central emphasis was the integration of rock-physics models to link elastic and petrophysical variables in a physically interpretable manner. \citet{grana2016bayesian} proposed a Bayesian linearised rock-physics inversion that provides an analytical solution under Gaussian priors, offering an efficient alternative to iterative nonlinear optimisation. Building on this idea, \citet{defigueiredo2017bayesian} developed a joint Bayesian framework that simultaneously estimates acoustic impedance, porosity, and lithofacies by coupling rock-physics priors with post-stack seismic data, while facies-based inversion incorporated lithology and fluid classes as discrete prior states \citep{zabihinaeini2017quantitative}.

	Uncertainty quantification also gained traction. Stochastic approaches such as MCMC for nonlinear elastic-impedance inversion \citep{zhang2015seismic} and geostatistical inversion that samples posterior petro-elastic properties \citep{bordignon2016integration,carmo2017exploring} moved beyond single deterministic solutions. Meanwhile, prior-model construction advanced through sedimentary-guided a priori models that embed seismic stratigraphy and depositional-pattern information, improving impedance inversion under sparse or biased well control \citep{zhao2016improve,kieu2015incorporating}.

	Machine-learning techniques appeared only sporadically, mostly as shallow models, including support vector machines for lithology prediction \citep{sebtosheikh2015lithology}, genetic-algorithm-optimised neural networks for porosity estimation \citep{kuroda2016analysis}, and neural-network mappings from seismic attributes to reservoir properties \citep{muradov2017application}. Hidden Markov models accounted for vertical lithological correlations \citep{feng2017reservoir}, and \citet{cao2017time} introduced machine learning for time-lapse property-change estimation, suggesting potential complementarity with physics-based 4D analysis. Overall, these applications remained isolated, and the dominant framework was still physics-driven inversion augmented by Bayesian and geostatistical uncertainty characterisation, forming the physical backbone for later data-driven advances.

	\subsubsection{2018--2019: Early adoption of deep learning}

	From 2018 to 2019, seismic-property research began shifting from incremental refinements of classical inversion toward end-to-end deep-network approaches, marking the initial adoption of deep learning for subsurface property interpretation \citep{das2019convolutional,biswas2019prestack}. Several studies showed that deep networks can learn direct mappings from seismic data to subsurface properties, reducing manual intervention and the computational cost of iterative inversion. For example, \citet{yang2019deep}, the most cited property study in the corpus, trained a supervised deep fully convolutional network to build velocity models directly from seismic data, bypassing the iterative tomography or full-waveform-inversion (FWI) workflow and its heavy reliance on human quality control, while a CNN trained on synthetic full-waveform seismograms generated from rock-physics-constrained earth models inverted normal-incidence data for elastic impedance and used approximate Bayesian computation to quantify posterior uncertainty \citep{das2019convolutional}. Deep learning was also applied to lithology prediction \citep{zhang2018deep}, and sequence-oriented architectures such as RNNs and TCNs were used to regress impedance and other petrophysical properties from seismic traces by exploiting their sequential structure \citep{alfarraj2018petrophysical,mustafa2019estimation}.

	A more consequential change occurred in training paradigms, which expanded beyond purely supervised learning toward physics-guided and weakly supervised frameworks. \citet{biswas2019prestack} proposed a physics-guided CNN that embeds the convolutional forward model into training, enabling prestack and poststack inversion without relying exclusively on pointwise elastic-property labels. Physics-informed neural networks further incorporated wave-equation constraints directly into the loss, improving physical plausibility and reducing the need for large labelled datasets \citep{xu2019physics}. Semi-supervised and adversarial strategies addressed label scarcity: \citet{alfarraj2019semi} used limited well-log supervision with learned forward operators and consistency constraints for impedance inversion, while GAN-based approaches were explored for rapid seismic-to-velocity mapping and data-driven velocity model building \citep{mosser2018rapid,arayapolo2019deep}.

	Meanwhile, physics-based and stochastic inversion continued to progress. Transdimensional MCMC was extended to quasi-3D impedance inversion with rigorous uncertainty quantification \citep{cho2018quasi}, and geostatistical inversion incorporated multiple uncertainty sources and was accelerated using distributed deep-learning frameworks \citep{pereira2019strategies,liu2019accelerating}. Sparse-representation methods and ensemble learners provided additional alternatives for multi-parameter inversion and large-scale lithologic mapping \citep{she2019data,kuhn2018lithologic}. Overall, end-to-end deep inversion, physics-guided learning, and label-efficient strategies collectively established deep learning as a viable, and often superior, complement to classical iterative inversion \citep{piresdelima2019convolutional,gao2019optimized}.

	\subsubsection{2020--2023: Deep-learning consolidation}

	Between 2020 and 2023, seismic-property research entered a consolidation stage of deep learning, with attribute modelling increasingly posed as an end-to-end mapping from seismic records to subsurface parameters such as impedance, velocity, porosity, and lithology \citep{li2020deep,wu2020seismica,adler2021deep}. Survey-style studies began to systematise neural-network inversion methods, suggesting that relatively stable technical and evaluation frameworks had emerged \citep{li2023comprehensive,grana2022probabilistic}.

	CNN-based inversion remained dominant, evolving from fully convolutional residual networks with transfer learning to multi-parameter prestack inversion \citep{wu2020seismica,das2020petrophysical}. Attention mechanisms and stronger temporal-modelling components, including TCNs, CNN--LSTM hybrids, and spatiotemporal modules, were introduced to capture nonlinear relationships and long-range dependencies along traces \citep{wei2021seismic,smith2022robust,mustafa2020spatiotemporal}, while residual attention further improved feature discrimination for impedance estimation \citep{wu2022seismic}. Toward the later part of this period, Transformer and CNN--Transformer hybrids appeared, leveraging self-attention to model global context \citep{fu2023seismic}. Multi-task formulations that jointly recover multiple parameters using shared encoders or uncertainty-weighted losses improved stability and noise robustness relative to single-parameter regression \citep{wu2021deep,zhao2021fluid}.

	In parallel, physics-guided closed-loop training advanced: network-predicted properties were passed through differentiable forward modelling to generate synthetic seismic data, and physics-consistency losses reduced reliance on dense labels \citep{sun2021physics,wang2020well,zhang2021robust}. PINNs, which embed wave-equation constraints into training, gained traction for velocity and impedance inversion under few-shot or label-free settings \citep{zhu2021general,liu2023joint}. Domain-knowledge-guided frameworks incorporated rock-physics constraints and geological priors to improve physical plausibility, and physics-aware stochastic inversion combined neural surrogates with probabilistic sampling for uncertainty-aware inference \citep{zhang2023domain,su2023seismic,brkle2023deep}.

	Transfer learning and label-efficient training were widely adopted to address limited samples and survey variability \citep{wu2020seismica,park2020automatic}. Semi-supervised and adversarial frameworks used sparse well-log supervision with consistency constraints or GAN-based learning to stabilise estimation under scarce labels \citep{wu2021semi,chen2022seismic}, while cycle-consistent GAN variants exploited unlabelled data and domain-adaptation strategies reduced the synthetic-to-field gap for cross-domain prediction \citep{zhong2020inversion,wang2022seismicb}. Weakly supervised multi-dimensional inversion enabled 1D well logs to supervise 2D/3D networks by masking losses outside well locations, reducing trace-by-trace discontinuities \citep{wu2021deep}. Additional efforts included autoencoder-based dimensionality reduction for large-scale inversion and RGT-constrained property interpretation to improve stratigraphic consistency \citep{gao2021large,di2022relative}.

	Uncertainty quantification received increasing, though still limited, attention, spanning Bayesian probabilistic inversion, uncertainty-propagation networks, and Monte Carlo dropout or ensembles within physics-guided training \citep{grana2022probabilistic,sun2021physics}. Overall, progress covered architecture design, learning-paradigm innovation, and deeper integration of physical and geological constraints \citep{wang2020velocity,zhang2022comparison}. Nevertheless, challenges persisted, including weakened structural features, limited cross-survey generalisation, strong dependence on well-log availability, and the lack of rigorous, calibrated uncertainty quantification for most deep-learning inversion methods.

	\subsubsection{2024--2025: Cross-survey generalization and emerging foundation models}

	In the 2024--2025 period, seismic-property inversion shifted toward cross-survey generalisation, generative modelling, and model universality \citep{li2024probabilistic,dou2024contrasinver,hosseinzadeh2025seismic}. Advances emphasised richer geological and physical priors, label-efficient learning under ultra-sparse well control, and multi-task joint-modelling. Most notably, foundation-model ideas emerged, aiming for unified, prompt-driven property interpretation cross-surveys.

	A major thread improved generalisation and geological plausibility by injecting explicit priors. New inversion targets with stronger rock-physics grounding, such as Poro-Acoustic Impedance (PAI), were proposed to better couple elastic responses with petrophysical meaning \citep{leisi2024poro}. Wavelet-transform-based inversion addressed low-frequency deficiencies and improved thin-bed resolution \citep{fu2024wavelet}. Other works compressed multi-attribute inputs via PCA for probabilistic porosity regression \citep{verma2025seismic} or introduced stratigraphic positional encoding in Transformer models to enforce depositional ordering \citep{zhou2025gamma}. Physics-guided frameworks matured further, including probabilistic PINNs that combine rock-physics forward models with Bayesian inference \citep{li2024probabilistic}, and model--data dual-driven networks that integrate physical constraints with data-driven learning for elastic and fracture-property inversion \citep{zhang2025joint,li2025seismicb}.

	The transition from CNN-dominated local modelling to global self-attention accelerated. Hybrid Transformer--CNN architectures were developed for poststack impedance inversion and prestack multi-parameter estimation \citep{ning2024transformer,yang2025hctnet}, while self-attention-based 3D frameworks such as TransInver explicitly leveraged long-range spatial correlations beyond fully convolutional designs \citep{li2024transinver}. Joint estimation of multiple parameters (e.g., impedance and porosity) within a single model further highlighted cross-parameter complementarity for improved identifiability \citep{sun2024seismic}.

	Label scarcity remained a central constraint. ContrasInver demonstrated contrastive semi-supervised inversion with as few as two wells \citep{dou2024contrasinver}, and domain-adversarial semi-supervised frameworks extended 1D supervision to 2D/3D settings through feature alignment \citep{zou2024domain}. Cycle-consistent GAN inversion continued to be refined for field applications \citep{fernandes2024cycle}. Meanwhile, diffusion models emerged for impedance inversion, enabling stable training and principled uncertainty sampling without the mode-collapse issues typical of GANs \citep{liao2025inverdiff,song2024seismic}. Bayesian modelling also diversified, including normalising-flow-based posterior approximations and Bayesian 4D inversion for time-lapse characterisation, while GPU-accelerated stochastic MCMC indicated continued practicality of classical probabilistic methods \citep{arabpour2025bayesian,kjnsberg2024bayesian,moon2024stochastic}. In parallel, prompt-driven and pretraining-based approaches began to explore unified modelling that treats seismic and well logs as conditional inputs, aiming at transfer across surveys and property types within a single representation \citep{yimindou2025geological,sheng2025seismic}.

	\subsubsection{Summary and task-specific outlook}

	At present, most machine-learning and AI-based studies on property interpretation remain highly focused on specific survey areas and specific target properties. The most common targets are impedance, velocity, and related attributes that are strongly linked to seismic reflections and admit relatively explicit physical mappings to reflection responses, which makes it easier to achieve strong performance on these quantities. Porosity occupies a middle ground: its physical relationship with acoustic impedance is well established through rock-physics models, but multi-factor coupling with lithology, fluid content, and pressure complicates purely data-driven prediction. In contrast, for properties that lack direct seismic-response mechanisms or are governed by stronger multi-factor coupling---such as gamma, lithology, and resistivity---data-driven methods often struggle to learn stable and transferable relationships. Their predictions are more susceptible to noise, non-uniqueness, and inter-survey variability, leading to particularly limited generalisation.

	Property interpretation also exhibits clear survey specialisation in practice \citep{zhang2023domain,wu2021deep}. Models are typically trained on well and seismic data from a single area and are primarily intended for that same deployment context. Cross-survey transfer often requires retraining or introducing complex adaptation strategies \citep{meng2022seismic,zhong2020inversion}. Moreover, in structurally complex settings with dense faulting, intense folding, or rapid stratigraphic variations, purely local data fitting is often insufficient to ensure spatial continuity and geological plausibility \citep{wang2020well,zhang2021robust,di2022relative}. Practical applications therefore frequently require additional strong constraints---such as structural frameworks, stratigraphic consistency, or facies control---to obtain property volumes that better satisfy interpretation needs \citep{sun2021physics,grana2022probabilistic,kolbjrnsen2020bayesian}.

	More recently, prompt-driven and foundation-model approaches have begun to reshape this landscape \citep{yimindou2025geological,sheng2025seismic}. By training on broader data distributions, such models can, at inference time, take seismic data together with well constraints or related conditional information as inputs and directly output the corresponding property model, enabling a more end-to-end prediction workflow. This strategy reduces error accumulation arising from multi-stage coupling in conventional pipelines and provides a scalable starting point for unified modelling cross-survey areas and properties, although large-scale field validation remains at an early stage.

	An important dimension of property-interpretation research is the distinction between poststack (acoustic impedance, porosity from AI) and prestack (elastic parameters, multi-parameter joint estimation) formulations. Throughout 2015--2025, poststack inversion often served as the initial testbed for new deep-learning architectures, owing to its lower dimensionality, simpler physics, and wider data availability, before methods were extended to prestack settings. Prestack inversion, which recovers richer elastic information (Vp, Vs, density), was more closely tied to rock-physics integration and physics-guided frameworks. Velocity model building constituted a third, largely independent thread with its own data requirements and evaluation protocols; it experienced a particularly concentrated growth during 2018--2019 \citep{yang2019deep,mosser2018rapid,arayapolo2019deep}, followed by steady development through physics-informed and differentiable-programming approaches.

	Looking forward, several directions are likely to continue advancing. Stronger explicit constraints on complex structure and spatial consistency are needed to improve the geometric and semantic reliability of predictions. More systematic treatment of non-uniqueness and uncertainty is required---diffusion models, which emerged in 2024--2025 for impedance inversion \citep{liao2025inverdiff,song2024seismic}, offer a principled generative framework for posterior sampling that may complement or supersede GAN-based approaches. Foundation models operating on raw seismic inputs may further blur the traditional boundaries between poststack, prestack, and velocity inversion without explicit data-type distinctions. Validation under larger-scale and more representative 3D benchmarks and evaluation protocols is essential to rigorously assess cross-survey generalisation and to drive property modelling from survey-specific solutions toward a more general and transferable unified paradigm.

	\subsubsection{Illustrative example: property field results}
	\begin{figure*}[htb]
		\includegraphics[scale=0.5]{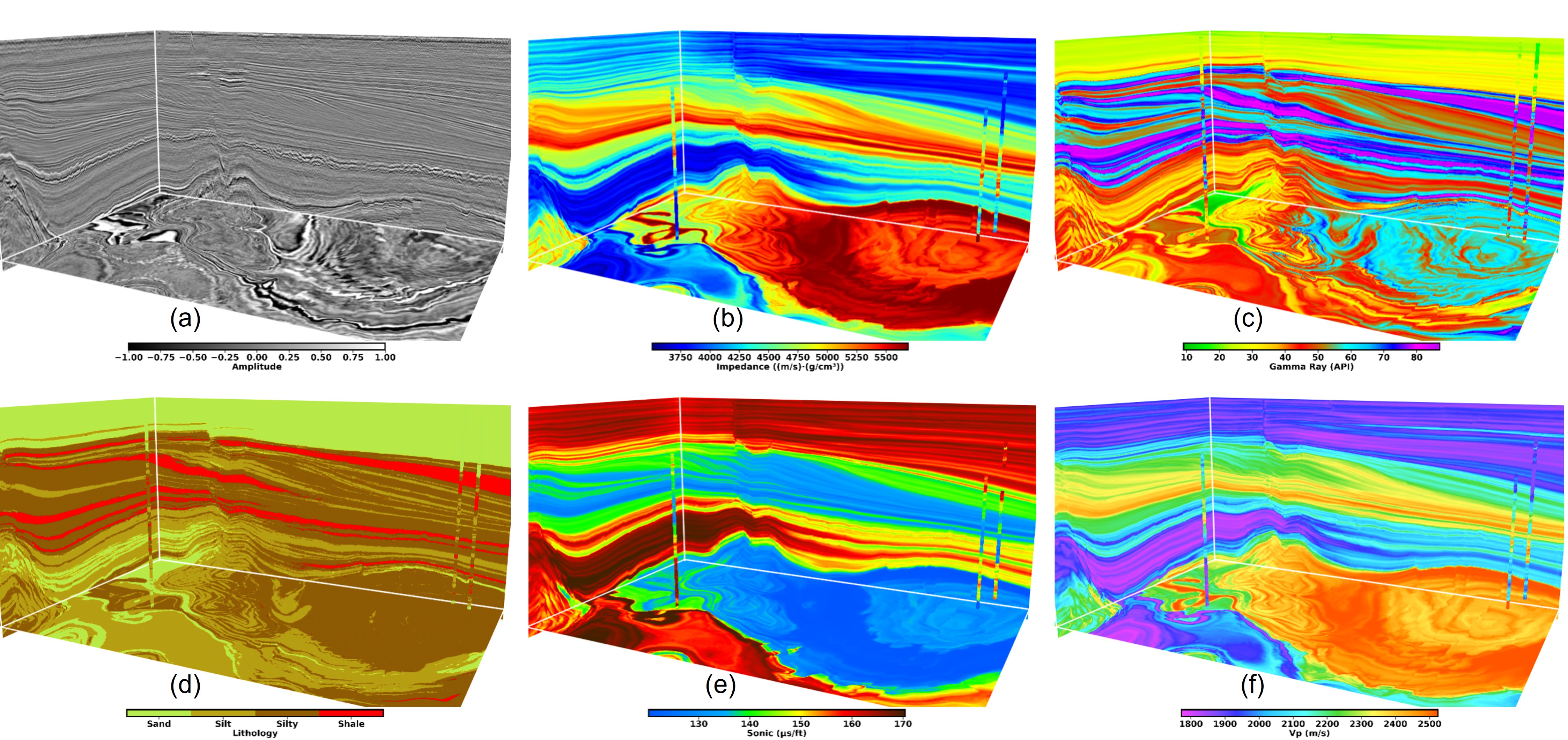}
		\centering\caption{Field outputs of the accompanying CIG-Bench property baseline, an HRNet backbone trained on synthetic dataset following the existing promptable conditional strategy of GEM \citep{yimindou2025geological}, taking seismic data and well logs as inputs. (a) Netherlands F3 seismic data. (b) Acoustic impedance. (c) Gamma-ray. (d) Lithology. (e) Sonic. (f) Vp. Shown only to illustrate qualitative behaviour; the synthetic quantitative evaluation is in the Supplementary Material.}
		\label{fig8}
	\end{figure*}

	Figure~\ref{fig8} illustrates the behaviour of the accompanying CIG-Bench property baseline, which under a unified promptable conditional strategy takes a seismic volume together with sparse well-log constraints and outputs dense 3D property volumes. The predicted properties show strong spatial consistency with seismic reflection patterns, preserve laterally continuous stratigraphic trends, and capture clear vertical and lateral variations across geological units; despite spatially sparse well inputs, the model produces coherent volume-wide property distributions. This example illustrates the potential of a unified, promptable seismic- and well-log-driven formulation discussed in the review above. Implementation details and the synthetic quantitative evaluation (including the effect of varying the number of conditioning well logs) are reported in the Supplementary Material (Sections~\ref{si:api}--\ref{si:quant}).

	The accompanying baselines were evaluated quantitatively on held-out synthetic data, since subsurface interpretation lacks an objective field ground truth. The full evaluation protocol, metric definitions, and the cross-task comparison against widely used open-source models (FaultSegv1/v2, FaultNet, FaultSSL, DeepRGT, ChannelSeg, KarstSeg) are provided in the Supplementary Material (Sections~\ref{si:setup}--\ref{si:quant}). Consistent with the scope stated there, those numbers measure relative ranking and the methodological upper bound on held-out synthetic data, not field performance, and are intended as a reproducible reference rather than a definitive performance ceiling.

	\section{Cross-task Connections and Co-evolution}\label{sec:crosstask}

	The four tasks reviewed above are usually studied in isolation, yet in both the interpretation workflow and the methods that drive them they are tightly coupled; treating them as a single system rather than four parallel problems is central to the perspective of this review.

	In the workflow, structural interpretation supplies the geometric and temporal scaffold---fault networks together with an RGT/horizon framework---within which every other task operates. Geobody and facies interpretation act as intermediate semantic layers that populate this scaffold, delineating depositional bodies and reflection units, and in turn provide priors, such as lithologic assemblages, depositional environments, and body geometries, for property inversion. The coupling is bidirectional rather than a one-way pipeline: property volumes recovered by inversion are routinely post-processed---through thresholding, clustering, or connectivity analysis---back into geobodies, while structural and facies constraints are increasingly injected into property networks to enforce spatial continuity and geological plausibility in complex settings. Errors propagate along this chain, so the reliability of a downstream property model is ultimately bounded by the quality of the upstream structural and facies interpretation.

	Methodologically, the tasks have advanced less as independent lines than as a shared front. The synthetic-data-training paradigm that broke the 3D-annotation bottleneck emerged first in structural interpretation with FaultSeg3D \citep{wu2019faultseg} and diffused rapidly to geobodies---SaltSeg \citep{shi2019saltseg} and ChannelSeg3D \citep{gao2021channelseg}---and to RGT estimation \citep{bi2021deep}, carrying with it a common encoder--decoder/U-Net backbone. Successive innovations, including class-balanced and boundary-aware losses, weak- and semi-supervised training under sparse labels, attention and transformer modules, and dropout- or Bayesian-based uncertainty, were typically pioneered on one task and transferred to the others within a year or two. The clearest sign of convergence is the recent move toward unified, promptable foundation models that fold structure, geobody, facies, and property estimation into a single pretrained representation \citep{yimindou2025geological,sheng2025seismic}, replacing four task-specific pipelines with one conditional model.

	A global view of the field's most influential works reinforces this reading. The studies that gained the widest recognition are, with few exceptions, those that introduced a transferable formulation rather than a task-specific refinement: recasting interpretation as CNN-based image segmentation or regression \citep{waldeland2018convolutional}, replacing iterative physical inversion with end-to-end deep inversion \citep{yang2019deep,das2019convolutional}, generating synthetic training data to escape the 3D-annotation bottleneck (Section~\ref{sec:taskreview}), and establishing shared benchmarks and evaluation practice \citep{alaudah2019machine}. Precisely because each proposed an idea that generalized beyond its original task, these works were taken up, and cited, across multiple task communities and now act as the connective hubs of the field. The corpus-level citation structure (Supplementary Material, Section~\ref{si:biblio}) is consistent with this picture: although each community still cites predominantly within its own domain---a quantitative signature of the very fragmentation this review seeks to address---the cross-domain links that do exist concentrate on this small set of paradigm-defining works. Impact has therefore tracked cross-task generality rather than within-task optimization, which is itself an argument for studying, and benchmarking, the four tasks jointly.

	The shared trajectory also implies shared limits. The three challenges of Section~\ref{sec:challenges}---interpretation under complex geology, cross-survey generalization under low information density, and the absence of reliable benchmarks---are common to all four tasks but bite with different severity: structural labels are geometrically well defined, whereas facies labels are semantically ambiguous and scale dependent, so the same technique generalizes unevenly across tasks. Progress on any single task is therefore increasingly gated by advances shared across all of them, which is the central argument for reviewing, and benchmarking, the four together.

	\section{Outlook}\label{sec:outlook}

	\begin{figure*}[htb]
		\includegraphics[scale=0.56]{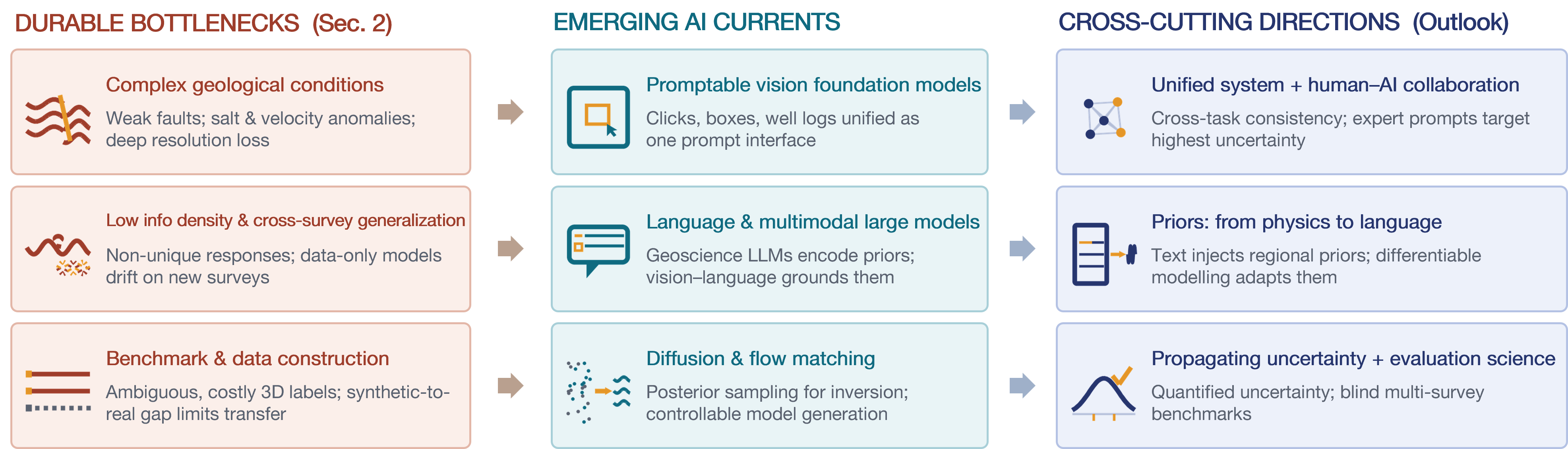}
		\centering\caption{Mapping the field's three durable bottlenecks
			(Section~\ref{sec:challenges}) to the emerging AI paradigms entering
			subsurface imaging understanding and to the cross-cutting research
			directions of this outlook. Each row links one bottleneck to the
			technical current that most directly addresses it and to the resulting
			research agenda.}
		\label{fig:outlook}
	\end{figure*}

	The task-wise reviews and the cross-task analysis above converge on a simple
	diagnosis: the durable bottlenecks of the field---interpretation under complex
	geological conditions, cross-survey semantic generalization under low
	information density, and the absence of reliable benchmarks
	(Section~\ref{sec:challenges})---are shared across all four task categories,
	and the innovations that have historically mattered most are precisely those
	that transferred across tasks rather than optimizing within one
	(Section~\ref{sec:crosstask}). Over the past two years, three technical
	currents from general-purpose AI---promptable vision foundation models,
	language and multimodal large models, and generative modelling based on
	diffusion and flow matching---have begun to enter subsurface imaging
	understanding, and each offers a promising, though still largely unproven, route toward one of these bottlenecks
	(Figure~\ref{fig:outlook}). We therefore organize the outlook not as a list
	of per-task extrapolations, which are given at the end of each task review,
	but as a set of cross-cutting
	directions in which we expect, and argue for, the next decade of progress,
	discussing explicitly the role of these emerging model paradigms. Where the
	evidence permits, we state our own position.

	\subsection{A unified interpretation system}

	The clearest lesson of the corpus is that impact has tracked cross-task
	generality: synthetic-data training, encoder--decoder segmentation, end-to-end
	deep inversion, and shared benchmarks each originated in one task and
	propagated to the others within a year or two (Section~\ref{sec:crosstask}).
	We expect the same logic to govern the next paradigm-level advance, and we
	therefore argue that it will not take the form of a fourth generation of
	single-task networks. Instead, the decisive step is joint modelling of the
	interpretation system itself: architectures and training objectives in which
	fault networks, the RGT/horizon framework, geobody and facies semantics, and
	property volumes are predicted under explicit cross-task consistency
	constraints---faults must displace the RGT field they cut, facies belts must
	respect the stratigraphic scaffold, and property distributions must honour
	both.

	A concrete and, to date, underexplored example is the explicit coupling of
	structural information with property inversion. The review shows that RGT
	constraints have been used in preliminary form to enhance stratigraphic
	consistency in facies classification and property interpretation
	\citep{di2021relative,di2022relative}, and that well-conditioned, promptable
	inversion has emerged \citep{yimindou2025geological,dou2024contrasinver}.
	Existing approaches, however, typically incorporate such structural information only as soft guidance, for example concatenating fault or RGT/stratigraphic attributes as extra input channels or weak regularizers that bias the prediction rather than enforce it. To our knowledge, no work goes further to inject the full products of
	structural interpretation---a fault-probability volume together with an
	RGT/stratigraphic framework---into a property-inversion network as an
	explicit, differentiable end-to-end consistency constraint, while requiring
	in turn that the inverted properties remain compatible with the structural
	interpretation (for instance, that property discontinuities coincide with
	fault surfaces and that property layering conforms to RGT isochrons). The
	technical obstacles have largely disappeared: promptable conditional
	architectures \citep{gao2026foundation,yimindou2025geological} and
	differentiable RGT formulations \citep{bi2021deep,geng2020deep} are both in
	place; what is missing is simply connecting them and defining the joint
	training objective. We regard such structure-conditioned inversion as the most
	natural first step from ``shared backbones'' toward ``enforced consistency,''
	and as an immediately actionable open problem identified by this review.

	Early unified and promptable models point toward system-level modelling
	\citep{yimindou2025geological,sheng2025seismic,gao2026foundation}, but they
	currently share a backbone rather than enforcing inter-task consistency;
	making that consistency an explicit, differentiable part of training remains
	open. Because errors propagate along the structure--facies--property chain
	(Section~\ref{sec:crosstask}), joint formulations are also the only principled
	route to controlling downstream reliability rather than merely measuring it
	after the fact.

	\subsection{Priors: from physics to language}

	Across all four tasks, the settings where purely data-driven models fail are
	exactly those where strong priors exist but are not used: fault geometry and
	displacement kinematics, stratigraphic topology and superposition, rock-physics
	relationships, and wave-equation or imaging-operator physics. The three
	established injection routes---prior-driven synthetic data generation,
	physical operators embedded in architectures, and constraint terms in losses
	\citep{wu2023sensing,sun2021physics}---have been validated largely in
	isolation and mostly on single tasks. Two extensions appear both feasible and
	consequential. First, relational and topological constraints (fault--horizon
	truncation rules, stratigraphic ordering, geobody connectivity) are still
	almost absent from training objectives, even though they are precisely the
	criteria by which human interpreters judge geological plausibility; RGT-based
	formulations \citep{bi2021deep,geng2020deep,di2022relative} provide a natural
	differentiable substrate for encoding them. Second, differentiable geological
	forward modelling---extending the synthetic-data paradigm of
	\citet{wu2019faultseg} from a static data generator into a trainable
	component---would allow the data-generating priors themselves to be adapted to
	a target basin, directly attacking the synthetic-to-real gap that currently
	caps cross-survey generalization.

	Beyond these two routes, vision--language models are opening a third
	channel of prior injection: geological knowledge encoded in natural language.
	Regional geological memoirs, lithofacies and depositional-environment
	descriptions, well-log interpretation reports, and textbook-level knowledge of
	structural styles carry a wealth of priors that, until now, could be used only
	implicitly by human interpreters. The vision--language alignment paradigm
	\citep{radford2021learning} demonstrates that image representations can share
	an embedding space with textual semantics; geoscience large language models
	such as K2 and GeoGalactica \citep{deng2024k2,lin2024geogalactica} show that
	geological knowledge in the domain literature can be encoded into model
	parameters at scale; and multimodal instruction-following models in remote
	sensing \citep{kuckreja2024geochat} show how such capabilities can be grounded
	on concrete Earth-observation imagery. Transferring this line to subsurface
	imaging would make text-conditioned interpretation possible: a single
	sentence such as ``en-echelon normal faults in a transtensional setting above
	a prograding deltaic package'' would inject regional structural and
	depositional priors into segmentation and inversion networks, without those
	priors having to be hand-translated into loss functions. Realizing this
	requires paired corpora of seismic images and geological text---abundant in
	interpretation reports, published figures, and their captions, yet never
	systematically assembled---and we regard their curation and alignment
	pretraining as a high-leverage, largely unoccupied direction for the coming
	years.

	\subsection{Propagating uncertainty}

	We regard uncertainty quantification not as one desirable output among many
	but as the defining requirement of a field whose inverse problem is
	fundamentally non-unique. The trajectory of the past decade---from dropout
	approximations \citep{feng2021uncertainty} through Bayesian and probabilistic
	inversion \citep{grana2022probabilistic} to diffusion-based posterior sampling
	\citep{liao2025inverdiff,song2024seismic}---shows the machinery maturing.
	Recent progress in generative modelling consolidates this route further:
	diffusion posterior sampling provides a principled framework for inverse
	problems with known forward operators \citep{chung2023diffusion}; diffusion
	priors have been used to regularize full-waveform inversion and improve its
	stability under complex conditions \citep{wang2023prior}; and controllable
	diffusion generation can synthesize velocity models conditioned on geological
	classes or well logs \citep{wang2024controllable}, which makes generative
	models simultaneously serve as controllable generators for the
	synthetic-data priors discussed above, unifying prior injection and
	uncertainty quantification within a single framework. These two uses---diffusion as a learned prior that regularizes a physics-driven inversion, and conditional diffusion that samples models directly from the data---outline the two complementary modes in which diffusion-based inversion is likely to develop. Their shared attraction is that repeated sampling returns not a single estimate but a family of data-consistent realizations whose spread is itself a posterior-based uncertainty estimate, obtained without the mode collapse that has long limited GAN-based inversion; and because the seismic forward operator is known and differentiable, the measurement likelihood can be folded into each denoising step, keeping every realization consistent with the recorded data rather than merely plausible. In practice this reframes classical regularized inversion as posterior sampling: the score network supplies the geological prior, the forward operator supplies the physics, and the reverse process explores the null space instead of collapsing it to one solution. What still blocks routine use is largely computational rather than conceptual: reverse diffusion over large 3D volumes is expensive, and conditioning is fragile when well control is extremely sparse. Flow matching
	\citep{lipman2023flow,liu2023rectified} replaces the stochastic denoising
	process with a deterministic ODE, reducing sampling steps by an order of
	magnitude with a simpler training objective---particularly attractive for the
	compute-sensitive setting of posterior sampling over large 3D volumes. Its
	introduction to subsurface interpretation and inversion is at present nearly
	blank, and we consider it a well-defined and well-timed direction.

	Three gaps nonetheless remain. Calibration is rarely verified: most reported
	uncertainties are relative confidence maps whose statistical meaning is
	untested, and label-ambiguity-aware evaluation \citep{guillon2020ground}
	remains the exception rather than the rule. Uncertainty is not yet
	interpretation-oriented: what interpretation requires is the probability of
	geologically meaningful, structured events---whether a fault is laterally
	continuous, whether two horizons are correctly correlated across a survey, or whether a geobody boundary lies at a given location, rather than voxel-wise variance. Most importantly, uncertainty does not yet
	propagate: although the review shows that structural interpretation bounds the
	reliability of every downstream product (Section~\ref{sec:crosstask}), no
	current workflow carries a posterior over fault networks or RGT fields into
	facies and property estimation. Joint generative modelling under a diffusion
	or flow-matching framework makes such end-to-end posterior propagation a
	technically attainable target for the first time: sampling the structural
	posterior and feeding each sample as a condition into the
	structure-conditioned inversion described above yields a property posterior
	marginalized over structural uncertainty. We consider this the single most
	valuable open problem identified by this review---and note that both of its
	building blocks, structure-conditioned inversion and generative posterior
	sampling, are now in place.

	\subsection{Human--AI collaboration}

	The task reviews independently converge on the same practical conclusion:
	fully automatic interpretation is not the near-term deployment target. For
	faults, dense small-scale systems make heavy interaction prohibitive while
	full automation remains unreliable; for geobodies and facies, semantic
	ambiguity makes purely automatic outputs difficult to trust; for properties,
	sparse well control is itself a form of interactive conditioning
	\citep{dou2024contrasinver,yimindou2025geological}.

	Notably, the technical vehicle for this collaborative paradigm has taken
	shape rapidly over the past two years, with a clearly traceable lineage: the
	general-purpose vision foundation model SAM established ``promptable
	segmentation'' as an interaction primitive \citep{kirillov2023segment};
	Seismic Fault SAM and FaciesSAM adapted point- and box-prompting to fault and
	facies segmentation \citep{chen2024seismica,atolagbe2025toward}; multimodal
	prompt engines extended interaction across surveys and geobody types
	\citep{gao2026foundation}; and GEM unified structural, geobody, and property
	tasks under a single promptable model \citep{yimindou2025geological}. This
	lineage suggests that ``prompting'' is evolving from an interaction technique
	into a unified conditioning interface: clicks, boxes, well logs, and---in
	combination with the language channel above---natural-language descriptions
	are different modalities of the same conditional pathway.

	We therefore expect the dominant operational paradigm to be automation with
	lightweight, targeted refinement: models produce full-volume initial
	interpretations together with calibrated uncertainty, and sparse human
	inputs---prompts, anchor picks, well ties---are treated as high-value
	supervision spent where uncertainty is highest. This reframes interaction
	design as an optimization problem (maximum reliability gain per unit of expert
	effort) and gives uncertainty quantification an immediate practical consumer.
	It also implies a research agenda that the field has barely begun: interfaces
	and training schemes in which corrections are not merely local edits but feed
	back into the model, so that expert knowledge accumulates across surveys
	rather than being spent anew on each one.

	\subsection{Toward an evaluation science}

	Reliable evaluation is currently the weakest link in the field's scientific
	infrastructure, and we argue it deserves to be treated as a research object in
	its own right. Synthetic benchmarks, including the resource accompanying this
	review, measure methodological upper bounds under a known distribution; field
	deployment is judged anecdotally. Closing this gap requires several
	coordinated developments. Blind, multi-survey held-out evaluation---in which
	test surveys are withheld from all participants, as pioneered by community
	challenges and open data efforts \citep{alaudah2019machine,chai2022open}---is
	the only credible measure of the cross-survey generalization that every task
	review identifies as the central obstacle. Metrics must become geologically
	aware: voxel-wise accuracy is insensitive to exactly the failures that matter
	in practice, such as broken fault connectivity, violated stratigraphic
	topology, or facies boundaries that are plausible but mislocated; boundary-,
	topology-, and connectivity-based measures, together with uncertainty-aware
	scoring, should become standard. For facies in particular, no benchmark can
	precede an operationally consistent task definition with hierarchical,
	reproducible labelling guidelines (Section~\ref{sec:taskreview}); we regard
	this standardization as a prerequisite, not an afterthought. Benchmarks must
	also be governed as evolving, versioned community resources rather than static
	releases---datasets, splits, metrics, and baselines updated under explicit
	versioning so that reported numbers remain comparable over time; the
	accompanying resource is intended as one step in this direction, not its
	conclusion.

	\subsection{Data ecosystems and scaling}

	Foundation-model results suggest that representations pretrained on large
	seismic corpora transfer under limited labels
	\citep{sheng2025seismic,chikhaoui2024self}, but the field's data ecosystem is
	not yet structured to sustain this route: much high-quality 3D data remains
	proprietary, metadata and preprocessing conventions are inconsistent, and the
	provenance of pretraining corpora is often undisclosed
	(Section~\ref{sec:limitations}). The central need is therefore the curation of \emph{AI-ready} data: raw seismic archives---heterogeneous in acquisition, processing, sampling, and amplitude conventions---must be converted into standardized, quality-controlled, and machine-consumable corpora with consistent geometry and normalization, documented preprocessing, and, where labels exist, harmonized annotation protocols. Without this curation layer, the scale that foundation models depend on is nominal rather than usable, since inconsistent conventions inject spurious cross-survey variability that no amount of parameters can absorb. Progress here is as much institutional as
	technical: standardized metadata and licensing frameworks for seismic data
	sharing, disclosure norms for pretraining corpora, and rights-preserving
	release mechanisms---for which controllable generative models
	\citep{wang2024controllable} already provide a ready tool in the form of
	synthetic surrogates of proprietary surveys---would compound the value of
	every methodological advance reviewed above. The multimodal direction
	discussed above also places a new demand on the data ecosystem: paired corpora
	of seismic images, geological text, and well logs are the prerequisite for
	text-conditioned interpretation, and their systematic curation is currently
	undertaken by no one.

	\subsection{Agents}

	LLM-based agents that translate natural-language goals into chained
	interpretation operations \citep{kanfar2025intelligent,ren2025seismology}
	offer a plausible integration layer for the collaborative paradigm sketched
	above, and the emergence of geoscience-specific language models
	\citep{deng2024k2,lin2024geogalactica} means this layer can carry domain
	knowledge rather than acting as a generic dispatcher. We caution, however,
	that agents inherit, rather than solve, every limitation of the underlying
	task models; their near-term value lies in orchestration and provenance
	tracking under human oversight, not in autonomous interpretation.

Where agents are most likely to matter first is the long, multi-stage nature of subsurface interpretation itself. A field study is rarely a single prediction; it is an extended pipeline---data loading and conditioning, denoising and attribute computation, fault and horizon extraction, RGT and framework building, geobody and facies delineation, property inversion, and finally cross-validation against wells---in which the output of each stage becomes the input of the next and errors accumulate along the chain (Section~\ref{sec:crosstask}). Today this pipeline is stitched together manually, and much of an interpreter's time is spent not on geological judgment but on moving data between tools, matching formats and coordinate conventions, and re-running steps when an upstream choice changes. An agent that plans and executes such a workflow---selecting the appropriate task model at each stage, propagating intermediate products and their uncertainties forward, and keeping a reproducible record of the decisions taken---would turn the fragmented, task-specific models reviewed here into a coherent interpretation system, addressing at the workflow level the same cross-task coupling that Section~\ref{sec:crosstask} argues for at the model level.

A closely related role is lowering the barrier imposed by professional interpretation software. Conventional platforms are powerful but notoriously complex, with steep learning curves, dense menus, and specialized parameter settings that make even routine operations costly and interpreter-dependent. Because these operations are ultimately scriptable, an agent that maps natural-language intent onto the underlying software and processing APIs could let interpreters describe what they want geologically and delegate the mechanics of execution, making expert workflows more accessible, less error-prone, and easier to reproduce. Realizing this will require exposing interpretation tools through well-documented, machine-callable interfaces and grounding the language layer in domain knowledge so that instructions are translated into geophysically sensible operations---one concrete motivation for the open, programmatically accessible baselines and APIs that accompany this review.

\section{Limitations}\label{sec:limitations}

Beyond the field-wide challenges discussed in Section~\ref{sec:challenges}, this
work has several limitations of its own that we state explicitly.

First, the in-corpus bibliometric overview (Supplementary Material, Section~\ref{si:biblio})
relies on Crossref metadata, whose coverage is not exhaustive, so it should be
read as an approximate snapshot of inter-task knowledge flow rather than an exact
measurement.

Second, the quantitative results of the accompanying CIG-Bench resource are
reported on synthetic data (Supplementary Material, Section~\ref{si:scope}). The reported
metrics thus characterize the methodological upper bound and the relative ranking
among methods on held-out data from the same distribution, rather than directly
estimating field performance; generalization to real surveys is supported in the
main text only qualitatively.

Third, the CIG-Bench baselines are deliberately lightweight, single-task models
intended as reproducible reference points rather than state-of-the-art methods.
They are not benchmarked head-to-head against large foundation models, since the
scale and heterogeneous, often undisclosed pretraining data of such models make
fair and reproducible comparison difficult.

Fourth, task and geobody coverage is not exhaustive: the accompanying resource
spans a representative but limited subset of interpretation tasks and target
types, and some categories are reviewed without being provided as unified
baselines.

Finally, several methods and datasets evaluated or re-used by the accompanying
resource were developed by the authors or members of their research group. To
mitigate the resulting bias, all baselines are evaluated under identical
preprocessing, data splits, and metric implementations, and the curated
literature metadata and citation matrix are released so that the analysis can be
independently reproduced or revised.

	\section{Conclusion}\label{sec:conclusion}

	This article reviews a decade of subsurface imaging interpretation, organized into four major task categories: structural interpretation, geobody identification, seismic facies analysis, and property estimation. Across all four, we trace how a decade of deep learning has shifted the field from fragmented, task-specific models toward unified and transferable frameworks, while highlighting that subsurface interpretation remains fundamentally different from other AI-driven imaging tasks: signals are blurry and non-unique, semantics are sparse, and densely reliable annotations are essentially unobtainable.

	Deep learning has fundamentally reshaped subsurface interpretation since 2018, driven by the convergence of scalable synthetic data, encoder--decoder architectures, and physics-guided training. Fault interpretation has progressed furthest toward deployment, yet persistent challenges remain in low-SNR intervals, complex tectonic styles, and cross-survey generalization. For other categories, the gap between research and operational reliability is substantially wider, constrained by label scarcity, semantic ambiguity, and the intrinsic non-uniqueness of seismic imaging. The open challenges and future directions distilled from this review are discussed in Section~\ref{sec:outlook}.

	To make these field-wide arguments concrete and reproducible, the review is accompanied by CIG-Bench, an open and evolving resource providing baseline models and qualitative field results for fault segmentation, RGT estimation, geobody segmentation, and property modeling, with its construction and quantitative evaluation documented in the Supplementary Material. We hope that this review, together with the accompanying resource, helps move subsurface imaging interpretation from experience-driven practice toward a more systematic and reproducible discipline.

	\section*{Data availability}
	
	The CIG-Bench synthetic datasets, the curated literature metadata table used for the bibliometric analysis (including paper-level category labels and the directed citation matrix), pretrained model checkpoints, and all configuration files needed to reproduce the results reported in this paper are released through the CIG-Bench project page at \url{https://douyimin.github.io/CIG-bench}. Datasets re-used from earlier CIG publications (the synthetic fault and stratigraphic volumes of \citet{wu2020building}, cigChannel \citep{wang2025cigchannel}, and the karst dataset of \citet{wu2020deepkast}) are redistributed in their original form together with the CIG-Bench-specific train/validation/test splits and unified preprocessing scripts.
	
	\section*{Code availability}
	
	The CIG-Bench open-source benchmark library, including the inference and deployment APIs, training scripts for all reported baselines, and the metric implementations used for the quantitative comparison reported in the Supplementary Material, is available at the CIG-Bench project page (\url{https://douyimin.github.io/CIG-bench}). Specific license terms for the datasets and code are provided in the repository.

	\section*{Acknowledgements}

	This work was supported by the National Natural Science Foundation of China (Young Scientists Fund, Category C; Grant No.~42504108), National Postdoctoral Program for Innovative Talents (Grant No.~BX20250436), the National Key Research and Development Program of China (Grant No.~2024YFC3012803-05), and the Natural Science Foundation of Anhui Province (Youth Project; Grant No.~2408085QD127).

	%
	%
%

	\bibliographystyle{naturemag-doi}

	\bibliography{literature}

	\clearpage
	\appendix
	\setcounter{section}{0}
	\setcounter{figure}{0}
	\setcounter{table}{0}
	\renewcommand{\thesection}{S\arabic{section}}
	\renewcommand{\thesubsection}{S\arabic{section}.\arabic{subsection}}
	\renewcommand{\thefigure}{S\arabic{figure}}
	\renewcommand{\thetable}{S\arabic{table}}
	\renewcommand{\theequation}{S\arabic{equation}}

	\makeatletter
	\let\CIG@oldaddcontentsline\addcontentsline
	\renewcommand{\addcontentsline}[3]{%
		\def\CIG@type{#2}\def\CIG@sub{subsection}%
		\ifx\CIG@type\CIG@sub\else\CIG@oldaddcontentsline{#1}{#2}{#3}\fi}
	\makeatother

	\begin{center}{\Large\bfseries Supplementary Material}\end{center}
	\vspace{0.5em}

	\section{Literature Analysis of the Corpus}\label{si:litpart}

	\subsection{Literature Collection and Corpus Curation}\label{si:corpus}

	The paper corpus was collected and curated as follows. We first built a literature retrieval and metadata acquisition pipeline based on Crossref. Using a set of keywords related to subsurface imaging and interpretation, we implemented scripts to automatically search and aggregate candidate papers published between 2015 and 2025, extracting basic information such as DOIs, titles, authors, abstracts, venues, and citation counts. The full keyword list (organised by the four task categories), the query date, and the post-retrieval filtering rules used to remove duplicates and clearly off-topic entries are released together with CIG-Bench so that the full retrieval-to-curation pipeline can be reproduced or extended by the community.

	Accordingly, the metadata and citation links used in this study were obtained from the Crossref database. This implies that the statistics may exhibit temporal lag, may be biased against venues with limited Crossref coverage (notably non-DOI SEG expanded abstracts and some regional journals), and some entries may be missing due to incomplete coverage or delayed updates; the citation-network view in Figure~\ref{fig1} should therefore be read as an approximate, in-corpus snapshot of inter-task knowledge flow rather than a universe-level measurement.

	On this basis, we initially partitioned the candidate papers into four major categories via keyword scanning. We then manually verified the automatic assignments paper by paper, with particular attention to misclassifications caused by ambiguous category boundaries, keyword polysemy, or insufficient abstract information. Labels were corrected and supplemented accordingly, improving classification accuracy and consistency while maintaining broad coverage. Because the four categories are conceptually overlapping---for example, a study that uses fault constraints to support impedance inversion is naturally relevant to both Structure and Property---a small fraction of papers were assigned to more than one category when their primary contributions clearly spanned multiple tasks. Consequently, the sum of category-wise paper counts reported in subsequent sections slightly exceeds the size of the deduplicated corpus ($652$). Within this corpus, $478$ papers explicitly adopt machine-learning or deep-learning methodologies, and these constitute the subset used for the temporal and methodological statistics in Figure~\ref{fig1-1}. The resulting corpus retains the scalability and reproducibility of automated retrieval, while reducing noise and systematic bias through human review, thereby providing a reliable foundation for subsequent survey analysis and citation relationship studies.

	We release the resulting metadata table together with CIG-Bench to facilitate community reproduction of the retrieval and categorization process, and to enable further extensions and analyses. Finally, because the automatically retrieved candidate set is large, citations in the main text are limited to the most relevant and representative works for the narrative of this review. Papers not cited should not be interpreted as lacking research value, but rather reflect space constraints and organizational choices inherent to survey writing.

	\subsection{Bibliometric Overview of the Curated Corpus}\label{si:biblio}

	This section provides a quantitative, in-corpus description of the curated
	literature. It is intended only as an approximate overview of the corpus
	assembled for this review and not as a free-standing bibliometric study.

\begin{figure*}[!htb]
		\includegraphics[scale=1.15]{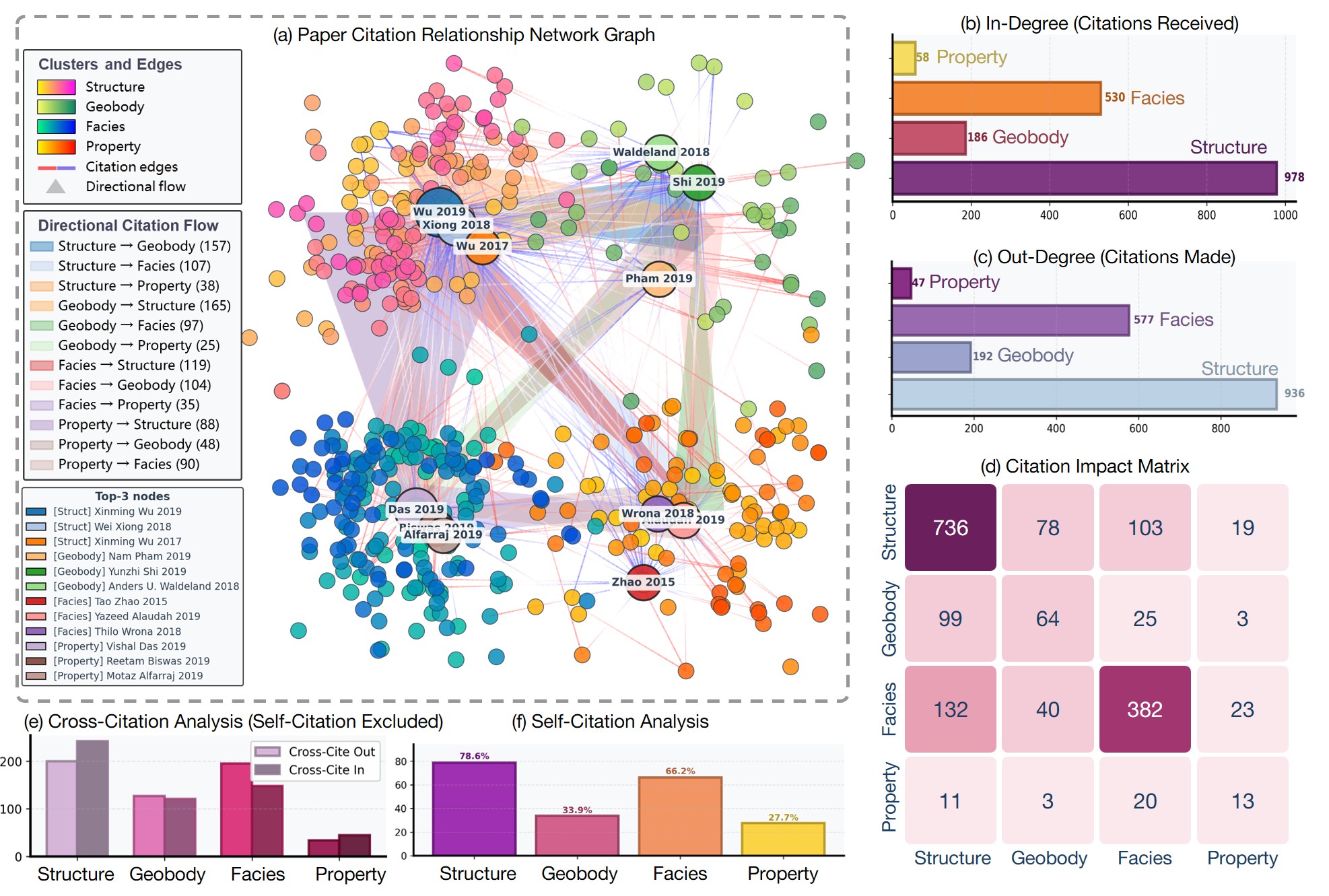}
		\centering\caption{Citation-network view of the curated corpus. From the 652 papers compiled in this work, panel (a) shows the paper-level citation network with cross-category linkages and a directional citation flow legend that aggregates inter-cluster citations under the manual category labels assigned during curation. Panels (b--f) summarise within-corpus directed citations between categories: (b,c) in-degree and out-degree counts per category; (d) the citation impact matrix; (e) cross-category citation flows after excluding within-domain citations, decomposed into cross-cite-out and cross-cite-in components; (f) the within-domain share of each category. The figure is provided as a quantitative overview of the curated corpus only.}
		\label{fig1}
	\end{figure*}

\begin{figure*}[htb]
		\includegraphics[scale=0.58]{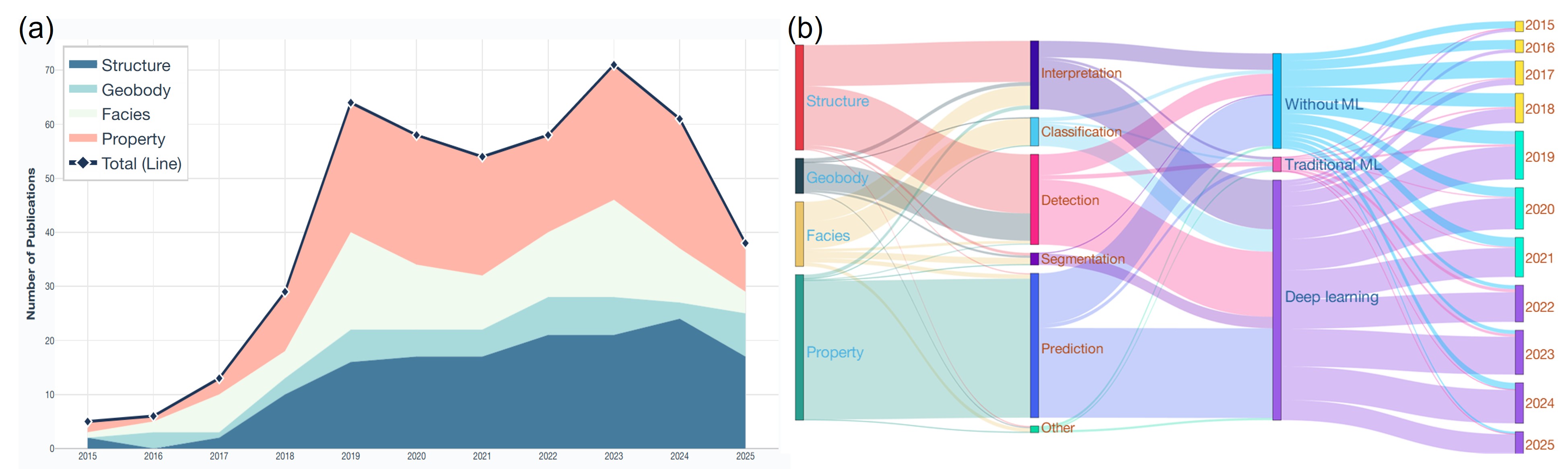}
		\centering\caption{Machine learning in subsurface imaging interpretation (2015--2025). (a) Stacked area chart of 478 ML papers across four targets (Structure, Geobody, Facies, Property), showing approximately tenfold growth from 2015 to a peak in 2023, with Property and Structure as leading categories. (b) Sankey diagram linking interpretation targets to tasks, methods, and publication years, revealing a decisive shift from traditional ML and non-ML approaches toward deep learning dominance post-2019.}
		\label{fig1-1}
	\end{figure*}

	Among the $478$ machine-learning-based publications used for the temporal analysis,\footnote{The full corpus comprises 652 deduplicated papers; the 478 subset reported here is the strict machine-learning/deep-learning subset on which the temporal and methodological statistics in Figure~\ref{fig1-1} are computed. The per-category paper counts in the citation-network view of Figure~\ref{fig1} (Structure 196, Geobody 72, Facies 122, Property 268) include some papers assigned to multiple categories and therefore sum to slightly more than 652.} Property and Structure dominate the literature, accounting for approximately 35.8\% and 31.6\%, respectively, followed by Facies at about 21.5\% and Geobody at about 10.7\%.

	This distribution reflects shared priorities in both research and practice. Property modeling typically serves high value applications such as reservoir characterization and quantitative prediction, generating direct benefits for reserve evaluation, development plan optimization, and risk control. It has therefore remained the most engineering driven and economically impactful direction, leading to the largest publication volume. Structure, in contrast, focuses on recovering fault systems and stratigraphic frameworks. Its outputs provide essential geometric constraints for horizon tracking, geobody boundary delineation, facies belt mapping, and the spatial consistency of property modeling, and often determine the upper bound of reliability for subsequent interpretation and modeling. It is thus a foundational component across multiple subsurface tasks and ranks second in publication volume.

	By comparison, Facies and Geobody more often function as intermediate representations that bridge seismic responses and geological semantics. They are typically used within an established structural framework to support depositional unit delineation, facies zoning, and object based characterization, and they provide priors for property modeling, including lithologic assemblages, depositional environments, and reservoir body geometries. In many industrial workflows, they may also appear as derived products obtained after property inversion and reservoir prediction, for example through thresholding, clustering, or connectivity analysis. Because label definitions for these tasks are more subjective and scale dependent, and are strongly influenced by local geological context, data quality, and imaging resolution, high-quality 3D annotation is costly and cross-survey generalization is difficult. These factors contribute to their comparatively smaller research scale and publication volume.

	Moreover, flow analysis in Figure~\ref{fig1-1}(b) indicates that deep learning became dominant rapidly after 2018 to 2019, while the share of traditional machine learning continued to shrink, although it remains present in certain classification and feature-engineering-oriented studies. Overall, Figure~\ref{fig1-1}(b) highlights a decade-long transition in subsurface interpretation from limited exploratory efforts to large-scale growth, accompanied by a structural methodological shift from traditional machine learning to deep-learning-dominated approaches.

	\section{The CIG-Bench Resource}\label{si:cigpart}

	\subsection{Scope and Design Rationale}\label{si:scope}

	The main text accompanies the review with an open, evolving benchmark resource,
	CIG-Bench, used only to make the review's arguments concrete and reproducible.
	Here we record the two design choices that govern how its results should be read.

	Although cross-survey generalization is the central open problem emphasized throughout the review, the quantitative results below are computed on synthetic data. This is a deliberate compromise: subsurface interpretation has no objective, absolutely correct ground truth, and densely annotated multi-survey field data essentially do not exist, so synthetic data are currently the only source that simultaneously provides exact voxel-level labels and controlled variation in structure, noise, and imaging conditions---at the known cost of the synthetic-to-real gap. Accordingly, the synthetic metrics in Section~\ref{si:quant} should be read as a controlled measure of the methodological upper bound and of the relative ranking among methods on held-out data from the training distribution, rather than as a direct estimate of field performance; generalization to real surveys is supported indirectly through the qualitative field examples shown in the main text. Establishing densely annotated multi-survey field benchmarks remains an important direction for future work.

	For the same reason, the baselines released here are deliberately lightweight and single-task. A benchmark must serve as a transparent and controllable reference point so that newly proposed methods can be compared under identical conditions. Compact, fully disclosed baselines fulfil this role: their architecture, hyperparameters, and training pipeline are fixed and reproducible, so any observed performance difference can be attributed to the candidate method rather than to opaque engineering or undisclosed implementation choices.

	\subsection{Dataset Details}\label{si:dataset}

\begin{figure*}[!htb]
		\includegraphics[scale=0.52]{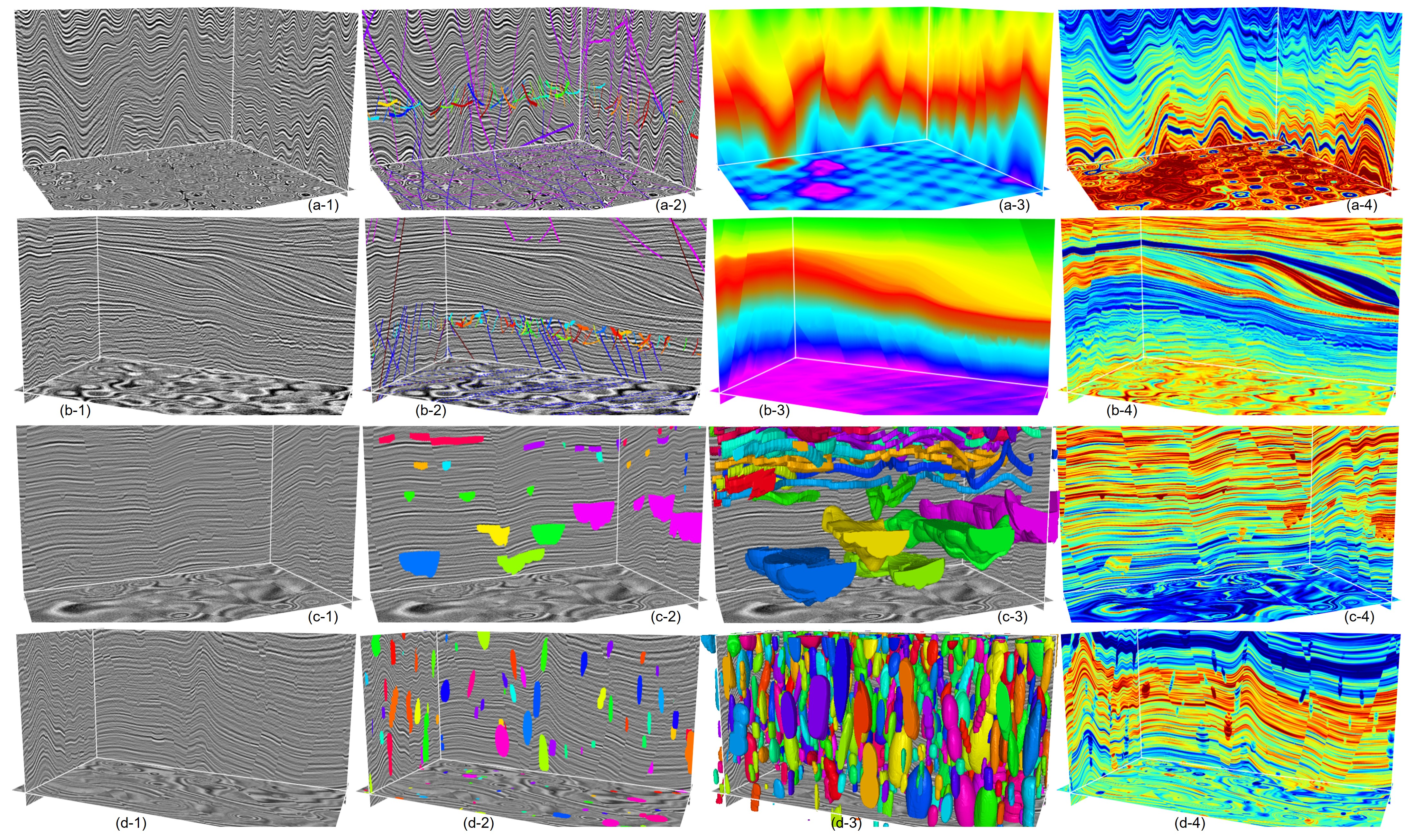}
		\centering\caption{Representative samples from the CIG-Bench synthetic dataset, arranged in four columns.
			Rows (a) and (b) show samples with structural annotations: raw seismic amplitude volumes (a-1,~b-1), fault labels with diverse fault styles (a-2,~b-2), relative geologic time (RGT) volumes describing the continuous depositional framework (a-3,~b-3), and the paired acoustic impedance/velocity volumes (a-4,~b-4).
			Rows (c) and (d) show samples with geobody annotations---channel systems (c) and karst cave systems (d) with different scales, geometries, and spatial distributions---displayed as raw seismic (c-1,~d-1), labelled geobodies (c-2,~d-2), three-dimensional geobody renderings (c-3,~d-3), and the paired acoustic impedance/velocity volumes (c-4,~d-4).  }
		\label{benchdata}
	\end{figure*}

	CIG-Bench provides a comprehensive collection of synthetic seismic volumes with paired labels spanning multiple geological structures and geobodies, including faults, stratigraphic sequences (represented as relative geologic time), velocity and acoustic impedance, fluvial channels, and karst cave systems. Each synthetic seismic volume has a dimension of $512 \times 1024 \times 512$ samples along the depth, inline, and crossline directions, respectively, with a vertical sampling interval of $5\,\text{m}$ (covering a depth range of approximately $0$--$2.56\,\text{km}$) and a lateral grid spacing of $25\,\text{m} \times 25\,\text{m}$ (corresponding to a surface survey area of approximately $25.6 \times 12.8\,\text{km}^2$, or roughly $328\,\text{km}^2$). Representative examples of these data and their corresponding labels are illustrated in Figure~\ref{benchdata}.

	As illustrated in Figure~\ref{benchdata}, rows (a) and (b) present seismic data samples annotated with fault and stratigraphic labels \citep{wu2020building}. Each sample set provides the raw seismic amplitude volume, a fault label volume encompassing diverse fracture styles, and a relative geologic time volume delineating the continuous stratigraphic framework. The fault labels cover representative structural styles including normal faults, reverse faults, multi-phase intersecting faults, and complex fracture assemblages, while the stratigraphic labels characterize structural features such as folds, unconformities, tilted strata, and complex depositional interfaces. Rows (c) and (d) present seismic data samples annotated with geobody labels, comprising typical reservoir geobodies such as channels \citep{wang2025cigchannel} and karst caves \citep{wu2020deepkast}, depicting targets of varying scales, morphologies, and spatial distributions within seismic responses that closely approximate realistic depositional settings. All samples are accompanied by ground-truth labels of acoustic impedance and velocity, enabling the construction of quantitative petrophysical models consistent with the simulated geological background, thereby supporting unified benchmarking of multiple tasks---including structural interpretation, geobody identification, relative geologic time estimation, and reservoir property inversion---under controlled and geologically reasonable conditions.

	\subsection{Baseline Implementations and Inference APIs}\label{si:api}

	This section records the inference APIs of the released baselines, referenced
	from the qualitative examples in the main text.

	\paragraph{Fault.} The fault model uses the skip-connection HRNet \citep{wang2021hrnet} variant and predicts a probability volume that is thresholded into a fault mask. Anisotropic rescaling (\texttt{scale\_t}, \texttt{scale\_h}, \texttt{scale\_w}) is exposed so that the model can be adapted to surveys with non-square spatial sampling. The optional \texttt{rank} and \texttt{chunk\_size} arguments split inference into chunks along the depth axis to keep GPU memory bounded on large field volumes \citep{li2025infer}.

	\begin{lstlisting}[style=cigpy]
		from cig_bench.predictor.fault import FaultPredictor

		fault_predictor = FaultPredictor(device="cuda")
		prob, used = fault_predictor.predict(
		seis,
		rank=4, chunk_size=64,    # memory-bounded inference
		threshold=0.5,
		scale_t=0.5, scale_h=0.85, scale_w=0.85,
		resize_back=True,         # return result at the original (T,H,W)
		)
		fault.visualize(used, prob)
	\end{lstlisting}

	\paragraph{RGT.} The RGT model regresses a smooth relative-geologic-time volume; horizons are then extracted as iso-surfaces of that volume. Optional sparse horizon annotations can be supplied as two auxiliary channels (\texttt{horizon\_rgt} and \texttt{horizon\_mask}) to constrain the prediction. The number of horizons returned by \texttt{extract\_horizons} is controlled by \texttt{n\_horizons}.

	\begin{lstlisting}[style=cigpy]
		from cig_bench.predictor.rgt import RGTPredictor

		rgt_predictor = RGTPredictor(device="cuda")
		rgt_vol, used = rgt_predictor.predict(seis)
		horizons = rgt_predictor.extract_horizons(rgt_vol, n_horizons=100)
		rgt_predictor.visualize(used, rgt_vol, horizons)
	\end{lstlisting}

	\paragraph{Geobody (channel/karst).} Both geobody predictors share the same multi-scale ensemble strategy: inference is run at multiple spatial scales by default (the specific scale set is configurable) and the resulting probability volumes are accumulated. A configurable post-processing step removes small connected components, which suppresses spurious responses in field data. The karst predictor is used identically; only the checkpoint changes.

	\begin{lstlisting}[style=cigpy]
		from cig_bench.predictor.channel import ChannelPredictor

		channel_predictor = ChannelPredictor(device="cuda")
		scores, used = channel_predictor.predict(
		seis,
		scales=[0.5, 0.75, 1.0, 1.25, 1.5],   # custom scale set
		accumulate="sum",
		)
		mask = channel_predictor.postprocess(scores, threshold=0.75, min_size=50000)
		channel_predictor.visualize(used, scores, mask)
	\end{lstlisting}

	\paragraph{Property.} The property predictor follows a promptable conditional paradigm: it takes a seismic volume together with a sparse well-log property volume (zeros where no well is present) and outputs a dense 3D property volume. Internally it stacks three channels --- seismic, sparse property, and a binary well mask --- and feeds them to the HRNet backbone \citep{wang2021hrnet}. The number and location of wells are not fixed; passing more wells generally improves accuracy, as quantified in Table~\ref{tab:cig_bench}(e).

	\begin{lstlisting}[style=cigpy]
		import numpy as np
		from cig_bench.predictor.property import PropertyPredictor

		prop_predictor = PropertyPredictor(device="cuda")
		vp_vol, used, wells = prop_predictor.predict(
		seis, vp_log,
		infer_shape=(640, 512, 512),
		)
		prop_predictor.visualize(used, vp_vol, wells)
	\end{lstlisting}

	\subsection{Training and Evaluation Setup}\label{si:setup}

	All CIG-Bench baselines are trained on the synthetic CIG-Bench training set described in Section~\ref{si:dataset} (or, for tasks that re-use earlier CIG datasets, on the corresponding training partitions of those datasets) and evaluated on held-out synthetic test volumes drawn from the same generation pipeline but never seen during training. Unless stated otherwise, all models are trained with the Adam optimizer on a single NVIDIA A100 GPU; per-task hyperparameters (learning-rate schedule, batch size, training iterations, patch size, and augmentation) are released together with the code and configuration files in the CIG-Bench repository to ensure full reproducibility. For the external baselines (FaultSegv1, FaultNet, FaultSegv2, FaultSSL, DeepRGT, ChannelSeg, KarstSeg), we use the authors' publicly released checkpoints and re-evaluate them on the CIG-Bench test sets under a unified preprocessing and metric implementation, so that the reported numbers reflect methodological differences rather than discrepancies introduced by inconsistent preprocessing, sampling strategies, or metric implementations. All metrics are computed at the volume level on the same test split per task, using a single-seed model; we do not currently report multi-seed variance or formal significance tests, and the reported numbers are intended as a reproducible reference rather than as a definitive performance upper bound.

	\subsection{Quantitative Comparison}\label{si:quant}

	The metric set is tailored to the geometric or volumetric nature of each task. Fault segmentation is evaluated with the voxel-level intersection-over-union, $\mathrm{IoU}=|P\cap G|/|P\cup G|$, together with the boundary-aware $\mathrm{HD}_{95}$ (the $95$th percentile of the symmetric Hausdorff distance between predicted and reference boundary point sets~\citep{huttenlocher1993hausdorff}) and the ODS/OIS contour-detection scores~\citep{arbelaez2011contour}, all computed on 2D inline/crossline sections. Channel and karst segmentation use IoU together with precision and recall. RGT estimation reports volumetric SSIM~\citep{wang2004ssim}, PSNR, and MAE on the RGT field itself, while $\mathrm{HD}_{95}$, ODS, and OIS are computed on equally spaced iso-contours extracted from 2D RGT sections, so that horizon-level geometric accuracy is assessed under a unified protocol. Property modelling reports MAE, SSIM, PSNR, and the coefficient of determination $R^{2}=1-\sum_i (y_i-\hat{y}_i)^2 / \sum_i (y_i-\bar{y})^2$ on the recovered volume, with LPIPS~\citep{zhang2018lpips} on 2D sections to capture perceptual fidelity. All metrics are reported as single-seed estimates without multi-seed variance.

	\begin{table}[htbp]
		\centering
		\caption{Quantitative comparison on the CIG-Bench benchmark across five
			seismic interpretation tasks. $\uparrow$ indicates higher is better,
			$\downarrow$ indicates lower is better. Best results are in \textbf{bold}.
			These numbers measure relative ranking and the methodological upper bound
			on held-out synthetic data, not field performance.}
		\label{tab:cig_bench}
		\renewcommand{\arraystretch}{1.2}

		\begin{tabular*}{\linewidth}{@{\extracolsep{\fill}} l c c c c @{}}
			\toprule
			\multicolumn{5}{c}{\textbf{(a) Fault segmentation}}\\
			\midrule
			Method & IOU\,\up & HD95 (2D)\,\dn & ODS (2D)\,\up & OIS (2D)\,\up \\
			\midrule
			FaultSegv1                       & 0.5926          & 16.3122         & 0.7372          & 0.7647 \\
			FaultNet                         & 0.6465          & 13.1779         & 0.7600          & 0.7898 \\
			FaultSegv2                       & 0.6675          & 11.3397         & 0.7811          & 0.8004 \\
			FaultSSL                         & 0.6899          & 11.0879         & 0.7689          & 0.7950 \\
			\midrule
			CIG-Bench (ours)                  & 0.7234          & 7.6465          & 0.7859          & 0.8083 \\
			\textbf{CIG-Bench (skip connect)} & \textbf{0.7455} & \textbf{5.2223} & \textbf{0.7865} & \textbf{0.8133} \\
			\bottomrule
		\end{tabular*}

		\vspace{1.0em}

		{\small
			\begin{tabular*}{\linewidth}{@{\extracolsep{\fill}} l c c c c c c @{}}
				\toprule
				\multicolumn{7}{c}{\textbf{(b) RGT estimation (stratigraphic sequence)}}\\
				\midrule
				Method & SSIM\,\up & PSNR\,\up & MAE\,\dn
				& HD95 (horizon, 2D)\,\dn
				& ODS (horizon, 2D)\,\up
				& OIS (horizon, 2D)\,\up \\
				\midrule
				DeepRGT                  & 0.9801          & \textbf{38.89} & \textbf{0.01557} & 8.6612          & 0.7754          & 0.8031 \\
				\midrule
				\textbf{CIG-Bench (ours)}& \textbf{0.9855} & 38.67          & 0.01866          & \textbf{4.7321} & \textbf{0.8306} & \textbf{0.8512} \\
				\bottomrule
			\end{tabular*}
		}

		\vspace{1.0em}

		\begin{minipage}[t]{0.49\linewidth}
			\begin{tabular*}{\linewidth}{@{\extracolsep{\fill}} l c c c @{}}
				\toprule
				\multicolumn{4}{c}{\textbf{(c) Channel segmentation}}\\
				\midrule
				Method & IOU\,\up & Precision\,\up & Recall\,\up \\
				\midrule
				ChannelSeg                & 0.8075          & 0.8701          & 0.9173 \\
				\midrule
				\textbf{CIG-Bench (ours)} & \textbf{0.8566} & \textbf{0.9202} & \textbf{0.9253} \\
				\bottomrule
			\end{tabular*}
		\end{minipage}\hfill
		\begin{minipage}[t]{0.49\linewidth}
			\begin{tabular*}{\linewidth}{@{\extracolsep{\fill}} l c c c @{}}
				\toprule
				\multicolumn{4}{c}{\textbf{(d) Karst segmentation}}\\
				\midrule
				Method & IOU\,\up & Precision\,\up & Recall\,\up \\
				\midrule
				KarstSeg                  & 0.8607          & 0.9101          & \textbf{0.9395} \\
				\midrule
				\textbf{CIG-Bench (ours)} & \textbf{0.8783} & \textbf{0.9502} & 0.9214 \\
				\bottomrule
			\end{tabular*}
		\end{minipage}

		\vspace{1.0em}

		\begin{tabular*}{\linewidth}{@{\extracolsep{\fill}} l c c c c c @{}}
			\toprule
			\multicolumn{6}{c}{\textbf{(e) Property modelling (CIG-Bench, varying number of well logs)}}\\
			\midrule
			\# Well logs & MAE\,\dn & SSIM\,\up & LPIPS (2D)\,\dn & PSNR\,\up & $R^{2}$\,\up \\
			\midrule
			4  logs              & 0.1298          & 0.8455          & 0.1633          & 25.66          & 0.7531 \\
			9  logs              & 0.1055          & 0.8891          & 0.1463          & 27.19          & 0.8121 \\
			16 logs              & 0.0868          & 0.9021          & 0.1222          & 27.97          & 0.8531 \\
			\textbf{25 logs}     & \textbf{0.0706} & \textbf{0.9347} & \textbf{0.1079} & \textbf{29.55} & \textbf{0.8616} \\
			\bottomrule
		\end{tabular*}

	\end{table}

	Table~\ref{tab:cig_bench} reports the quantitative comparison of representative methods on the CIG-Bench test sets across five seismic interpretation tasks. For fault segmentation (Table~\ref{tab:cig_bench}a), the CIG-Bench baseline surpasses the four widely used open-source models (FaultSegv1~\citep{wu2019faultseg}, FaultNet~\citep{dou2022loss}, FaultSegv2~\citep{li2024faultseg}, FaultSSL~\citep{dou2024faultssl}) across all four metrics, and the skip-connection variant is best overall; the most pronounced gain is on the boundary-aware HD95, which drops by roughly two thirds relative to FaultSegv1. For RGT estimation (Table~\ref{tab:cig_bench}b), DeepRGT~\citep{bi2021deep} and the CIG-Bench baseline recover the volumetric RGT field with comparable voxel-wise fidelity, but the CIG-Bench baseline produces substantially sharper horizon-level geometry, halving HD95 and improving ODS/OIS on extracted iso-contours. For geobody segmentation (Tables~\ref{tab:cig_bench}c,d), the CIG-Bench baseline improves IoU and precision over ChannelSeg~\citep{gao2021channelseg,gao2020channel} on channels and over KarstSeg~\citep{wu2020deepkast,yan2025karst} on karst caves; the slightly lower karst recall reflects a more conservative prediction strategy, generally preferable where weak boundaries make false positives costly. For property modelling (Table~\ref{tab:cig_bench}e), all five metrics improve monotonically as the number of conditioning well logs increases from 4 to 25, confirming that the promptable conditional baseline effectively leverages additional well constraints.

	These readings should be interpreted in light of the scope clarified in Section~\ref{si:scope}: the metrics measure relative ranking and the methodological upper bound on held-out synthetic data, not field performance, and are intended as a reproducible reference rather than a definitive performance ceiling.

	\newpage
	\begin{appendices}

	\end{appendices}

\end{document}